\documentclass[aps,pre,showpacs,twocolumn,superscriptaddress,groupedaddress]{revtex4} 

\usepackage[english]{babel}
\usepackage{graphicx}
\usepackage{amsgen}
\usepackage{amsfonts}
\usepackage{amssymb}
\usepackage{amsbsy}
\usepackage{amsmath}
\usepackage{bbm}
\usepackage[breaklinks,pdfborder={0 0 0}]{hyperref}
\usepackage{tikz}
\usetikzlibrary{plotmarks}
\usepackage[strict]{changepage}

\begin{document}

\title{Modeling record-breaking stock prices}

\author{Gregor Wergen}
\affiliation{Institut f\"ur Theoretische Physik, Universit\"at zu K\"oln, 50937 K\"oln, Germany}

\begin{abstract} We study the statistics of record-breaking events in daily stock prices of 366 stocks from the Standard and Poors 500 stock index. Both the record events in the daily stock prices themselves and the records in the daily returns are discussed. In both cases we try to describe the record statistics of the stock data with simple theoretical models. The daily returns are compared to i.i.d.~RV's and the stock prices are modeled using a biased random walk, for which the record statistics are known. These models agree partly with the behavior of the stock data, but we also identify several interesting deviations. Most importantly, the number of records in the stocks appears to be systematically decreased in comparison with the random walk model. Considering the autoregressive AR(1) process, we can predict the record statistics of the daily stock prices more accurately. We also compare the stock data with simulations of the record statistics of the more complicated GARCH(1,1) model, which, in combination with the AR(1) model, gives the best agreement with the observational data. To better understand our findings, we discuss the survival and first-passage times of stock prices on certain intervals and analyze the correlations between the individual record events. After recapitulating some recent results for the record statistics of ensembles of $N$ stocks, we also present some new observations for the weekly distributions of record events.

\end{abstract}
\date{\today}
\maketitle

\tableofcontents

\section{Introduction}

Not only because of the recent financial crises, the study of extremes in stock markets is of great importance for scientists and economists \cite{Sornette2003,Fama1965,Longin1996,Johansen1998}. Traders are eagerly interested in the statistics of extreme events in stock prices at the worlds stock exchanges. In the context of extreme and dramatic developments in finance, people also talk a lot about record-breaking events. A record is an entry in a series of events that exceeds all previous values. At the stock markets, a record stock price is often considered to be an important and remarkable event that attracts more attention and media coverage than others. 

In recent years, the theory of records has found many applications in various areas of science. Most extensively studied was the statistics of record temperatures and their connection with global warming \cite{Redner2006,Meehl2009,Wergen2010,Elguindi2012,Rahmstorf2011,Wergen2012b}. In 2010, Wergen and Krug~\cite{Wergen2010} presented a simple analytical model that predicts the effect of climatic change on the occurrence of daily and monthly temperature records to a good accuracy (see also \cite{Wergen2012b,Rahmstorf2011}). Furthermore, the statistics of records found applications in evolutionary biology \cite{Krug2005}, physics \cite{Oliveira2005,Sibani2006,Sibani2007}, hydrology \cite{Vogel2001} and of course also in sports \cite{Gembris2002,Gembris2007}. Additionally, a lot of progress was made from the mathematical point of view. Often motivated by the multitude of applications, the theory of records from time-dependent and correlated random variables was developed further (see for instance \cite{Wergen2013}). The interesting problem of the record statistics of independent random variables with a linear drift is well understood by now \cite{Wergen2013,Ballerini1987,Franke2010}. Also the ramifications involved with discrete distributions and ties due to rounding were studied \cite{Wergen2013,Gouet2005,Wergen2012}. But most important for this work was the full characterization of the universal record statistics of symmetric random walks by Majumdar and Ziff \cite{Majumdar2008} in 2008.

\medskip
Recently, we started analyzing the statistics of record-breaking stock prices. Preliminary results were already published in 2011 in the context of a study of the record statistics of biased random walks \cite{Wergen2011b} (see also \cite{Bogner2009}) and, in 2012, in an analysis of ensembles of independent random walkers \cite{Wergen2012a}. The purpose of this article is to present a more thorough and comprehensive discussion of record-breaking events in stock prices and returns. In particular, we will discuss and explain the occurrence of new upper and lower records in stock data by comparing them to various stochastic models.

As in \cite{Wergen2011b}, we study record events in daily stock data from the Standard and Poors 500 (S\&P 500) stock index \cite{SPdata}. Our data set contains 5000 consecutive trading days from 366 stocks that stayed in the S\&P 500 for the entire time-span from January 1, 1990 to March 31, 2009. We are interested both in the record events in the time series of the stocks themselves and in record-breaking daily returns. In a series of stock prices $S_0,S_1,...,S_n$, we have a record at the $n$th day if 
\begin{eqnarray}
 S_n > \textrm{max}\{S_0,S_1,...,S_{n-1}\}.
\end{eqnarray}
Analogously, we have a record breaking return $\Delta_n:=S_n-S_{n-1}$, if 
\begin{eqnarray}
 \Delta_n > \textrm{max}\{\Delta_{0},\Delta_1...,\Delta_{n-1}\},
\end{eqnarray}
where we set $\Delta_0=0$. When we consider the time series of stock prices $S_i$ or returns $\Delta_i$, the most important quantity for us is the probability that a certain entry in such a series is a record. This probability $P_n$ for a stock price $S_n$ is defined as
\begin{eqnarray}
 P_n := \textrm{Prob}\left[S_n > \textrm{max}\{S_0,S_1...,S_{n-1}\}\right].
\end{eqnarray}
For the returns $\Delta_i$ we define the probability $p_n$ in the same manner:
\begin{eqnarray}
 p_n := \textrm{Prob}\left[\Delta_n > \textrm{max}\{\Delta_0,\Delta_1,...,\Delta_{n-1}\}\right].
\end{eqnarray}
In the following, we will also refer to these quantities as the \textit{record rates}. Of similar importance are the closely related record numbers of the stock prices $R_n$ and the returns $r_n$, the numbers of records that occur in a time series up to step $n$. One obtains the important mean record numbers $\langle R_n \rangle$ and $\langle r_n \rangle$ of the stocks and the returns by summing over the respective record rates:
\begin{eqnarray}
 \langle R_n\rangle := \sum_{k=0}^n P_k \quad\textrm{and}\quad \langle r_n\rangle := \sum_{k=0}^n p_k.
\end{eqnarray}
This article discusses the record rates and record numbers of daily stock prices and returns in the S\&P 500 and compares them with several simple stochastic models such as simple i.i.d.~RV's or biased random walks. The aim of this work is to better understand the occurrence of record-breaking events in the stock markets and to find useful and accurate models that reproduce and predict them correctly. 

\medskip
Since this work summarizes multiple observations and results, we will now give a short outline of the rest of this article: We start by briefly recapitulating some important classical results from the theory of records in time series of independent and identically distributed (i.i.d.) random variables (RV's) in section \ref{fin:iid_rvs}. There, we also present the findings for the symmetric random walk derived by Majumdar and Ziff \cite{Majumdar2008}. Thereafter we discuss the record statistics of biased random walks with a Gaussian jump distribution following the results derived in Wergen et al.~\cite{Wergen2011b} and Majumdar et al.~\cite{Majumdar2012}. Subsequently, in section \ref{fin:autoregressive}, we introduce the more complicated, so-called autoregressive AR(1) process, which might be able to describe the statistics of record-breaking stock prices more accurately. In recent years, this model and its continuous analog, the Ornstein-Uhlenbeck process, was used to model stock data by several research groups \cite{Barndorff2001,Cont2003}. We analyze and discuss its record statistics numerically.

As a second alternative to the simple random walk model we will also introduce the generalized autoregressive conditional heteroscedasticity (GARCH) model \cite{Bollerslev1986}, which is particularly often used to model financial time series \cite{Engle2001,Duan1995}. In this more complicated model a second stochastic process is employed that describes the time dependence of the variance of the increments. We will discuss the record statistics of a specific GARCH process employing numerical simulations. 

Section \ref{fin:single_stocks} is about the record statistics of individual stocks from the S\&P 500 index. After introducing the data and analyzing some of the basic statistical properties of the time series of stock prices, we first have a look at the record statistics of the daily returns. We compute the record rate and the mean record number of the daily returns and discuss the correlations between individual return records. Then the record statistics of the stock prices themselves are analyzed. Here, we first compare the stocks to the simple Geometric Random Walk model of stock prices, before we discuss the stock records in the context of the more complicated AR(1) model and the GARCH model. Following this, we also consider the full distribution of the record number of stock prices and compare this distribution with theoretical predictions. We will briefly discuss the first-passage statistics and survival probabilities of the stocks, since they are closely related to the statistics of records in random walks.

\begin{figure*}
\includegraphics[width=0.48\textwidth]{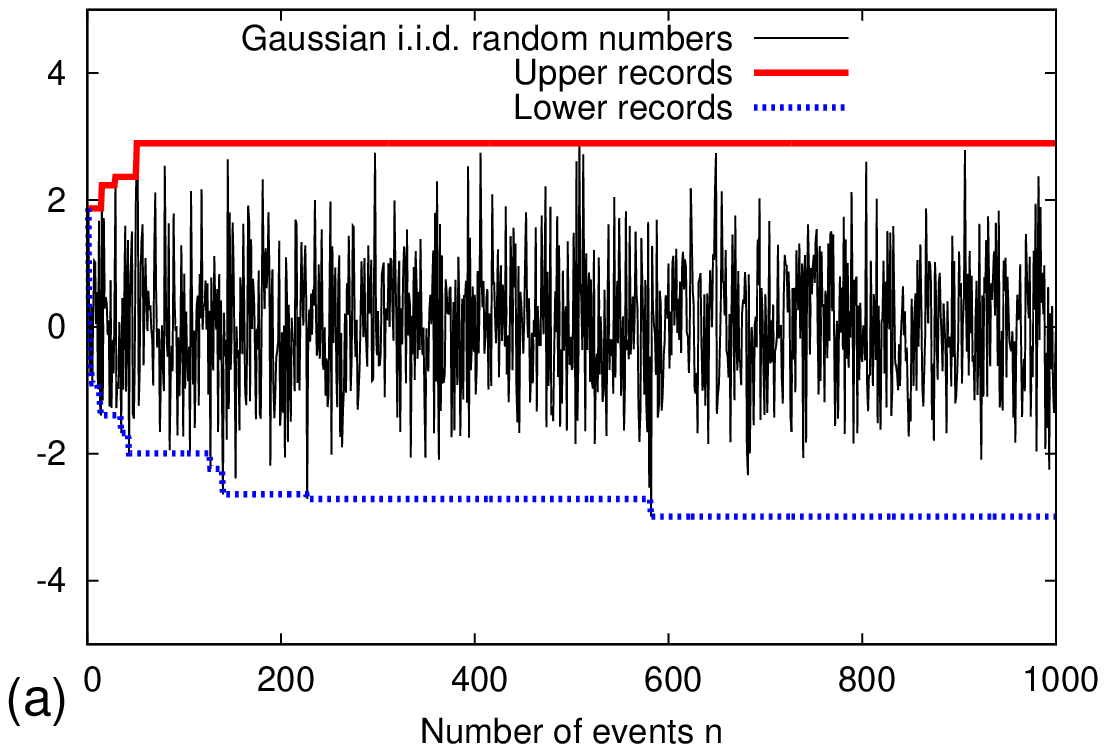}
\includegraphics[width=0.48\textwidth]{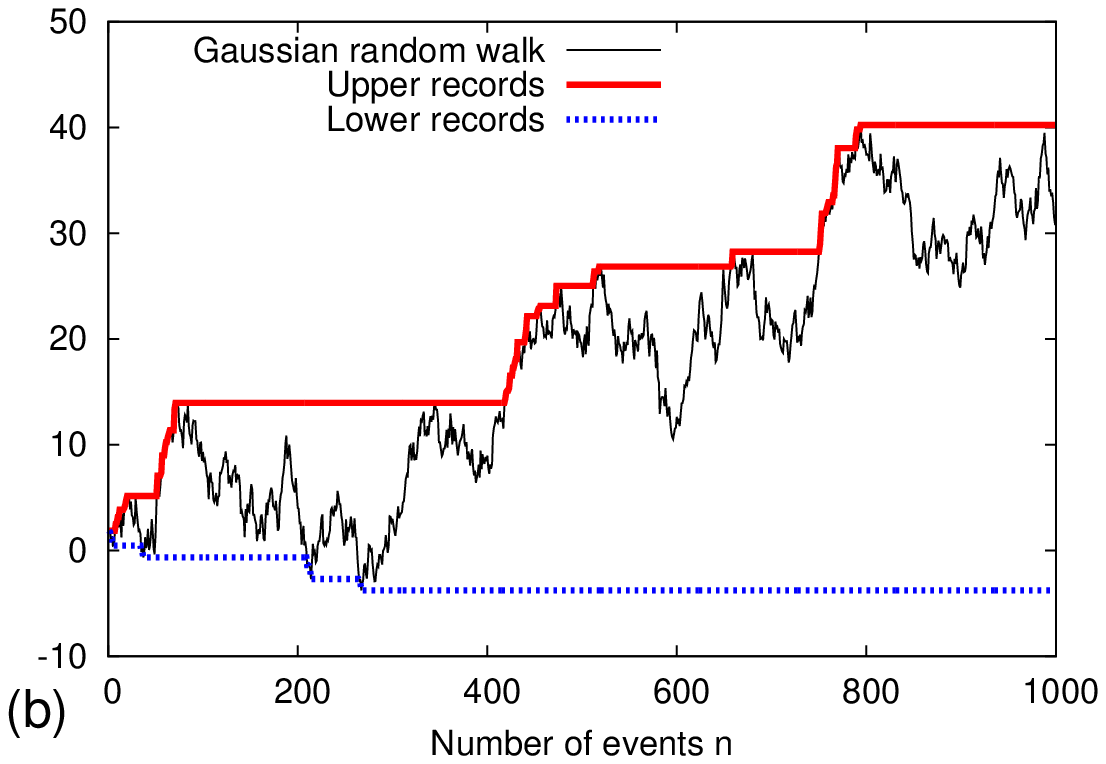}
\caption{\label{Fig:iid_rw_comp} (Color online) (a) 1000 numerically generated random numbers from a Gaussian standard normal distribution (with standard deviation $\sigma=1$). The thick red line gives the progression of the upper record, the dashed blue line the progression of the lower record. (b) A numerical realization of a random walk with 1000 steps. The jump distribution is again a Gaussian standard normal distribution with mean value zero and standard deviation unity. Again, the thick red and dashed blue lines give the progressions of the upper and lower record values respectively. }
\end{figure*}

In section \ref{fin:n_stocks}, we consider ensembles of stocks and compare their record statistics to the one of multiple independent random walks. The results in this section are mostly already published in \cite{Wergen2012a}, here, we present them also to make this work more comprehensive and self-contained. 

Some new findings for the weekly distribution of the record rate of the daily stock prices and the daily returns are presented in section \ref{fin:weekly_monthly}. These observations are partly in contradiction with the assumptions and findings discussed before and show that simple, random walk type models can only predict certain features of the stocks behavior, while others are not captured. 

Finally, in section \ref{fin:summary}, we will briefly summarize and evaluate our findings and discuss open question, which present possibilities for future research.

\section{Records in i.i.d.~RV's \& discrete-time random walks}
\label{fin:iid_rvs}

The classical theory of records from i.i.d.~RV's has been developed several decades ago. A detailed introduction can for instance be found in the books of Arnold et al.~\cite{Arnold1998} or Nevzorov \cite{Nevzorov2004} (see also \cite{Wergen2013}). For a series of i.i.d.~RV's $\xi_0,\xi_1...\xi_n$ sampled from a single probability density $f\left(\xi\right)$, we can easily give the probability for a certain entry in this series to be a record. Using the so-called \textit{stick-shuffling} argument one finds that, on average, the $\left(n+1\right)$th entry $\xi_n$ is a new record in $1/\left(n+1\right)$ cases. Therefore, the record rate of i.i.d.~RV's is given by
\begin{eqnarray}
 p^{\left(\textrm{iid}\right)}_n = \frac{1}{n+1}.
\end{eqnarray}
With this, it is also straightforward to compute the mean record number $\langle r_n\rangle$. By summing over the record rate we find that, for $n\gg1$, $\langle r_n\rangle$ grows logarithmically in $n$:
\begin{eqnarray}
 \langle r_n\rangle = \sum_{k=1}^n p_k = \sum_{k=1}^n \frac{1}{k+1} \approx \ln\left(n\right) + \gamma.
\end{eqnarray}
Here, $\gamma\approx0.577215...$ is the Euler-Mascheroni constant \cite{Arnold1998}. Interestingly, these results are completely independent from the choice of the underlying distribution $f\left(\xi\right)$. 

In 2008, Majumdar and Ziff \cite{Majumdar2008} found that the record statistics of a discrete-time random walk with a symmetric jump distribution has similar universal properties. They considered random walks with entries $X_0,X_1,...,X_n$ given by
\begin{eqnarray}\label{fin:symm_rw}
 X_i = X_{i-1} + \xi_i,
\end{eqnarray}
for $i\geq 1$ with i.i.d.~RV's $\xi_i$ sampled from a symmetric and continuous jump distribution $f\left(\xi\right)$. As an initial value they set $X_0=0$. The different characteristics of the record processes of i.i.d.~RV's and random walks are illustrated in Fig.~\ref{Fig:iid_rw_comp}. 

Majumdar and Ziff computed the full distribution of the record number $R_n$ of such a process. They obtained the probability of having $R_n$ records in a random walk of $n$ steps by subdividing the process into a series of $R_n$ first-passage and survival problems (cf.~\cite{RednerBook}). Those were then solved using a celebrated theorem by Sparre Andersen \cite{SparreAndersen1953,SparreAndersen1954}. Majumdar and Ziff showed that the probability $P_n$ for a record in the $n$th event of a symmetric random walk is the same as the survival probability $Q_n:= \textrm{Prob}\left[X_1,X_2,...,X_n>0\right]$, i.e. the probability that a random walk starting from the origin stays above the origin without crossing it for the next $n$ steps. According to Sparre Andersen \cite{SparreAndersen1953,SparreAndersen1954,Majumdar2008}, this probability and therefore also the record rate of the symmetric random walk is universal for symmetric random walks with a continuous jump distribution. They found that, for $n\gg1$, the record rate $P_n$ decays as
\begin{eqnarray}
 P_n \approx \frac{1}{\sqrt{\pi n}}.
\end{eqnarray}
This allows to compute the mean record number of the symmetric random walk:
\begin{eqnarray}
 \langle R_n\rangle = n \left( 2n \atop n\right) 2^{-2n+1} \approx \sqrt{\frac{4n}{\pi}}.
\end{eqnarray}
Here, the mean record number grows with $\sqrt{n}$, much faster than in the case of i.i.d.~RV's. The most important and surprising feature of these results is that, for arbitrary $n$, they are again completely independent from the choice of the symmetric and continuous jump distribution $f\left(\xi\right)$. They also hold for the so-called L\'evy flights, i.e. random walks with a jump distributions lacking a finite second moment.

\label{fin:random_walks}

More important for an application to financial data, as we will see in the following sections, are random walks with a bias. In the contexts of the evolution of stock prices, such a bias represents an inherent growth in the system, a long term interest-rate or economic growth. The entries $X_0,X_1,...,X_n$ of a discrete-time random walk with bias $c$ can be described with $X_0=0$ and 
\begin{eqnarray}
 X_i = X_{i-1} + \xi_i + c,
\end{eqnarray}
where the $\xi_i$'s are again sampled from a symmetric and continuous distribution $f\left(\xi\right)$. Unfortunately, the full universality, which was found for the symmetric random walk is not conserved here. In the biased case, the record statistics generally depends on the shape of the jump distribution $f\left(\xi\right)$.

In a recent publication of Majumdar et al.~\cite{Majumdar2012}, the asymptotic record statistics of such biased random walks were studied thoroughly. The authors used a generalized version of Sparre Andersen's theorem to compute the survival probabilities of biased random walks for different classes of jump distributions. With their findings, they could derive asymptotically exact results for the distribution of the record number, as well as the extreme value statistics of the ages of the shortest and longest lasting records.

The most relevant parameter for the asymptotic survival and record statistics of a biased random walk is the so-called L\'evy-index of the jump distribution $f\left(\xi\right)$. Majumdar et al.~considered jump distributions, whose Fourier transforms $\tilde{f}\left(k\right) := \int_{-\infty}^{\infty} f\left(\xi\right)e^{-ik\xi}\mathrm{d}\xi$ have the following small $k$ behavior:
\begin{eqnarray}
 \tilde{f}\left(k\right) \approx 1 - \left(\alpha_{\mu}|k|\right)^{\mu}.
\end{eqnarray}
Here, $\mu$ is called the L\'evy-index and $\alpha_{\mu}$ is a parameter that represents a characteristic length scale of the jump distribution. The L\'evy index describes the tail-behavior of the jump distribution (for $|\xi|\gg1$). For distributions with a finite second moment, like a Gaussian, an exponential or a uniform distribution, one always finds a L\'evy-index of $\mu=2$. For $n\rightarrow\infty$, these jump distributions lead to a random walk that behaves like classical Brownian motion with a mean square displacement $\langle \sum_{k=1}^n X_k^2 \rangle \propto n$ (cf.~\cite{Weiss1994}). 

More complicated is the regime of $0<\mu<2$, where the jump density has no finite variance. In this case, one finds that, for large $|\xi|\rightarrow\infty$, the real-space representation of the jump distribution is of the form $f\left(\xi\right) \propto |\xi|^{-1-\mu}$ and decays with a power law broader than $1/|\xi|^{3}$. These jump distributions are called \textit{heavy-tailed}, random walks with heavy-tailed jumps are called L\'evy flights.

Majumdar et al.~found that, for strongly heavy-tailed distributions with $\mu<1$, the drift does not change the asymptotic behavior of the mean record number $\langle R_n\rangle$. Here, we have $\langle R_n\rangle \propto \sqrt{n}$ independent of $c$. For the marginal case of $\mu=1$, which includes the well known Cauchy distribution (c.f. \cite{Abramowitz1970}), they found that the mean record number depends non-trivially on $c$. In this regime, for $n\rightarrow\infty$, one can show that $\langle R_n\rangle \propto n^{\Theta\left(c\right)}$ with $\Theta\left(c\right) = 1/2 + 1/\pi\;\textrm{arctan}\left(c\right)$. 

More interesting for us are the distributions, which are less heavy-tailed than the Cauchy distribution, with $\mu>1$. In this regime, for $c>0$, the mean record number grows linearly in $n$ and one finds $\langle R_n \rangle \approx \alpha_{\mu}\left(c\right) n$ with $\alpha_{\mu}\left(c\right)$ depending on the exact shape of $f\left(\xi\right)$. This asymptotic behavior of the mean record number was also found for the Brownian case with $\mu=2$ and $c>0$. However, the asymptotic distributions of the record number and the statistics of the extremal ages of the longest and shortest lasting record are systematically different between the $1<\mu<2$-regime and the the case of $\mu=2$ with a positive $c$.

In the context of our interest in stock prices, the Brownian regime with $\mu=2$ will prove to be of particular importance. For the special case of a Gaussian jump distribution with probability density 
\begin{eqnarray}
 f\left(\xi\right) = \frac{1}{\sqrt{2\pi\sigma^2}}e^{-\frac{\xi^2}{2\sigma^2}},
\end{eqnarray}
Majumdar et al.~\cite{Majumdar2012} derived an explicit expression for the prefactor $\alpha_2\left(c\right)$. Here, for $n\rightarrow\infty$, they found that
\begin{eqnarray}\label{fin:R_n_Gauss_asymp}
 \langle R_n \rangle \approx n\;\textrm{exp}\left[-\sum_{k=1}^{\infty} \frac{1}{2k} \textrm{erfc}\left(\frac{c\sqrt{k}}{\sigma \sqrt{2}}\right)\right].
\end{eqnarray}

\begin{figure*}
\includegraphics[width=0.48\textwidth]{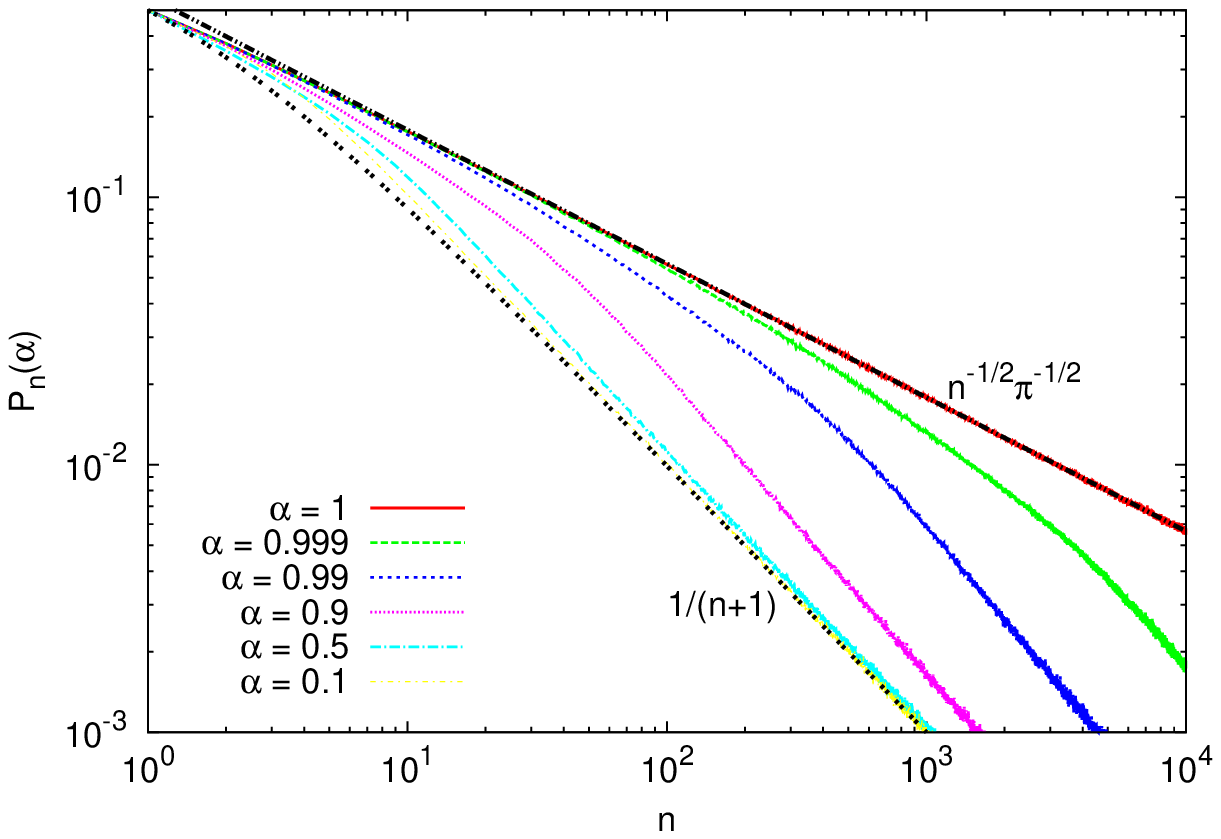}
\includegraphics[width=0.48\textwidth]{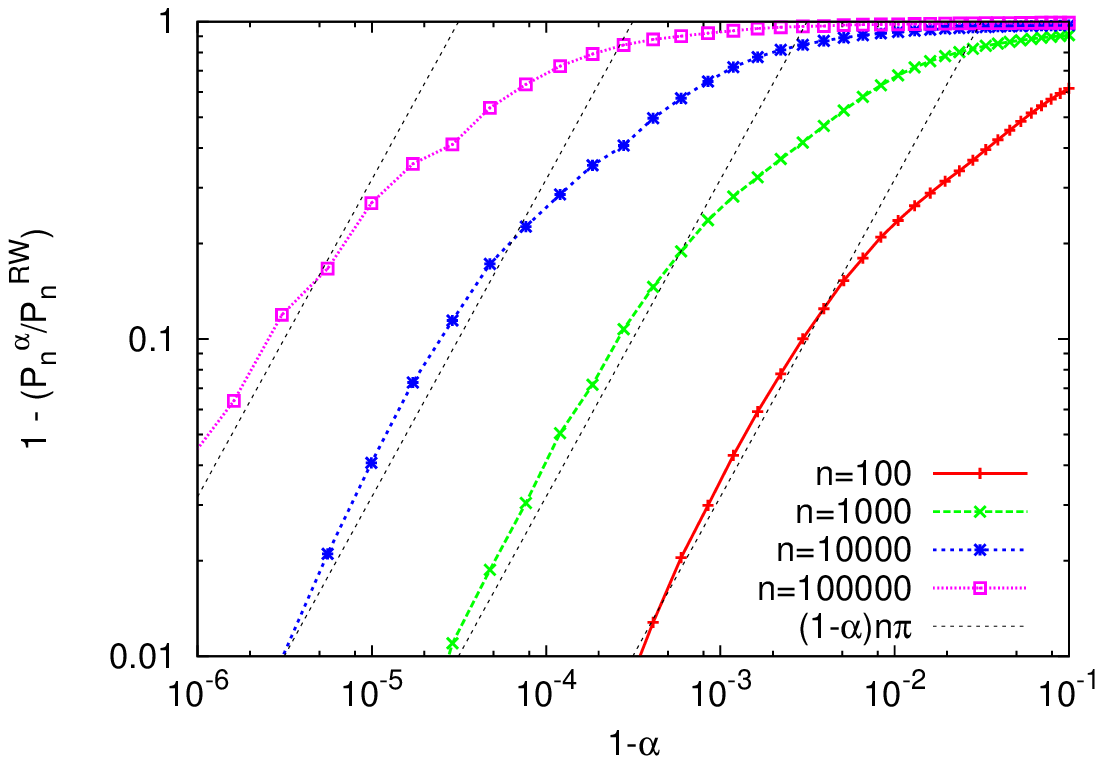}
\caption{\label{Fig:ar1_n10000} (Color online) (a) Numerical simulations of the record rate $P_n$ for the AR(1) process with different values of $\alpha$ and $n=10^4$ steps. The increments $\xi_i$ were sampled from a Gaussian jump distribution with $f\left(\xi\right) = \frac{1}{\sqrt{2\pi}}e^{-\xi^2/2}$. We averaged over $10^6$ realizations for each value of $\alpha$ and simulated $\alpha=1$ (the random walk case), $\alpha=0.999$, $\alpha=0.99$, $\alpha=0.9$, $\alpha=0.5$ and $\alpha=0.1$. We also plotted our analytical results for the random walk and i.i.d.~RV's (black dotted lines). For large values of $\alpha$ close to one the process behaves like a random walk for a long time. For smaller $\alpha$ he converges to the i.i.d.~behavior quite quickly. Eventually we expect $P_n\propto1/n$ for all $\alpha<1$ in the $n\rightarrow\infty$ limit. (b) Numerical simulations of the record rate $P_n^{\left(\alpha\right)}$ of the AR(1) process for different values of $n=10^2,10^3,10^4$ and $10^5$. We plotted the effect of $\alpha$ on $1-\left(P_n^{\left(\alpha\right)}/P_n^{\left(1\right)}\right)$ against small values of $1-\alpha$ between $10^{-6}$ and $10^{-1}$. We simulated $10^6$ time series for each value of $\alpha$ and $n$. For very small values of $1-\alpha$ ($\alpha$ very close to one), $1-\left(P_n^{\left(\alpha\right)}/P_n^{\left(1\right)}\right)$ is well described by $\frac{n}{\pi}\left(\alpha-1\right)$. This indicates that here the record rate of the AR(1) process is well described by $P_n^{\left(\alpha\right)} \approx P_n^{\left(1\right)} \left(1-\frac{n}{\pi}\left(1-\alpha\right)\right)$.}
\end{figure*}

Another regime, which has proven to be useful for the analysis of stock prices, was studied by Wergen et al.~\cite{Wergen2011b} in 2011. They considered the case of a small drift $c$ and a finite $n$ with $c\sqrt{n}\ll\sigma$ for random walks with a Gaussian jump distribution. Here, the generalized Sparre Andersen theorem for the survival probability is helpful too and one finds that, for $c\ll\sigma/\sqrt{n}$:
\begin{eqnarray}
P_n \approx \frac{1}{\sqrt{\pi n}} + \frac{c}{\sigma}\frac{\sqrt{2}}{\pi}\textrm{arctan}\left(\sqrt{n}\right).
\end{eqnarray}
For $n\gg1$ (but still $c\sqrt{n}\ll\sigma$) this leads to
\begin{eqnarray}\label{fin:Pred_finite_n}
 \langle R_n\rangle \approx \frac{2\sqrt{n}}{\sqrt{\pi}} + \frac{cn}{\sqrt{2}\sigma}\quad\textrm{and}\quad P_n \approx \frac{1}{\sqrt{\pi n}} + \frac{c}{\sqrt{2}\sigma}.
\end{eqnarray}
It is important to notice that these expressions are entirely linear in $c$. Because of this, the effect of the drift on the record rate and the mean record number of lower records can be obtained by simply replacing the '$+$'-signs in Eq.~\ref{fin:Pred_finite_n} by '$-$'-signs.

In \cite{Wergen2011b}, these results (Eq.~\ref{fin:Pred_finite_n}) were compared to numerical simulations of a biased Gaussian random walk and it was shown that they also apply for other jump distributions with an existing second moment. 

In section \ref{fin:single_stocks} we will discuss these findings for biased random walks, especially the ones for the Gaussian process given in Eqs. \ref{fin:R_n_Gauss_asymp} and \ref {fin:Pred_finite_n}, in the context of records in the evolution of stock prices. Prior to this, we will now introduce two more complicated processes, which might be even more accurate in the modeling of stock records, and analyze their record statistics numerically.

\section{Records in autoregressive processes}
\label{fin:autoregressive}

\subsection{Records in the AR(1) process}

The fact that the record statistics of i.i.d.~RV's as well as of symmetric random walks is well understood by now, and also our previous work on financial data has motivated us to consider another stochastic process, which, in some sense, intermediates between uncorrelated RV's and random walks. The natural way to interpolate between these two processes is a process of entries $X_0,X_1,...,X_n$ with $X_0=0$ and
\begin{eqnarray}\label{fin:ar1}
 X_i = \alpha X_{i-1} + \xi_i,
\end{eqnarray}
and $\alpha$ being a parameter between zero and one. The $\xi_i$'s are again i.i.d.~RV's sampled from a continuous and fully symmetric jump distribution $f\left(\xi\right)$. In the special case of $\alpha=0$, this process is just a series of i.i.d.~RV's with $X_1=\xi_1,X_2=\xi_2, ...$~. For $\alpha = 1$ we recover the symmetric random walk (cf. Eq.~\ref{fin:symm_rw}). An important feature of such a process is that he is generally not invariant under translations and the increments $X_i-X_{i-1}$ depend crucially on the value of $X_{i-1}$. Therefore the distance of the walker from the origin is relevant. Such a model is also known as an AR(1) process (a series, where $X_i$ depends in a autoregressive manner on the previous $k$ values $X_{i-1},...,X_{i-k}$ is generally called an AR($k$) process).

Unfortunately, the methods introduced in the previous publications concerning the record statistics of random walks do not allow to compute the record statistics of the AR(1) process. Here, Sparre Andersen's theorem loses its validity and, by now, it was not possible to derive the record rate and the mean record number using an alternative analytical approach.

In the left panel of Fig.~\ref{Fig:ar1_n10000} we show numerical results for the record rate of the AR(1) process for different values of $\alpha$ with a Gaussian jump distribution $f\left(\xi\right) = \frac{1}{\sqrt{2\pi}}e^{-\xi^2/2}$. Apparently, for small values of $\alpha$, this process behaves almost like a series of i.i.d.~RV's. If $\alpha$ is close to one the record statistics resembles the one of a symmetric random walk for a long time, but eventually the record rate decreases and approaches the $1/n$ behavior as well. These simulations allow us to conjecture the asymptotic behavior of the record rate of the AR(1) process. For $n\rightarrow\infty$, we assume that
\begin{eqnarray}
 P_n^{\left(\alpha\right)} \xrightarrow[]{n\rightarrow\infty} P_n^{\left(\textrm{iid}\right)} = \frac{1}{n}
\end{eqnarray}
In the opposite regime of finite $n$ and $\alpha\approx 1$, we expect that the record statistics of the process differs only slightly from the random walk. The mean record number will grow roughly proportional to $\sqrt{n}$ as in the case of the symmetric random walk. As indicated by several studies of the AR(1) process \cite{Novikov1993,Alili2005,Ditlevsen2007,Novikov2008} and our own numerical simulations, the survival probability of this process decays exponentially with $n$ (and not with $1/\sqrt{n}$ as in the case of the symmetric random walk). Building upon the work of Novikov \cite{Novikov2008}, we can make an educated guess for the record rate $P_n^{\left(\alpha\right)}$ for $\left(1-\alpha\right)\ll1$:
\begin{eqnarray}
 P_n^{\left(\alpha\right)} \approx P_n^{\left(1\right)} e^{-\left(1-\alpha\right) B_{n}} \approx P_n^{\left(1\right)}\left(1-\left(1-\alpha\right)\right) B_n) 
\end{eqnarray}
with a positive parameter $B_n$ depending only $n$ and $f\left(\xi\right)$, but not on $\alpha$. Interestingly, for the Gaussian AR(1), very good agreements with numerical results were found for a $B_n = \frac{n}{\pi}$ linear in $n$. For $\left(\alpha-1\right)\rightarrow0$ this would lead to
\begin{eqnarray}\label{fin:p_n_alpha_guess}
 P_n^{\left(\alpha\right)} \approx \frac{e^{-\frac{n}{\pi}\left(1-\alpha\right)}}{\sqrt{\pi n}} \quad\textrm{and}\quad\langle R_n^{\left(\alpha\right)}\rangle \approx \frac{2\sqrt{n}}{\pi} e^{-\frac{n}{\pi}\left(1-\alpha\right)}.
\end{eqnarray}
This conjecture is compared with numerical simulations of the record rate of the AR(1) process in the right panel of Fig. \ref{Fig:ar1_n10000}. Apparently, for very small values of $\left(1-\alpha\right)$, Eq. \ref{fin:p_n_alpha_guess} gives a very good approximation for $P_n^{\left(\alpha\right)}$.

Despite these observations, we will use numerical results for our analysis of the stock data. It would be an interesting and challenging goal for future research to compute the record statistics of the AR(1) process analytically. We also simulated the record rate of the AR(1) process for other jump distributions, but found only a very weak dependence of $P_n^{\left(\alpha\right)}$ on the shape and tail-behavior of $f\left(\xi\right)$. However, it seems as if the Gaussian AR(1) gives an upper limit for $P_n^{\alpha}$ and $\langle R_n^{\alpha}\rangle$. Alternative jump distributions (for instance heavy-tailed ones) can only decrease these quantities slightly.

In our analysis of stock data, it will be interesting to see if a possible autoregressive component with a parameter $\alpha<1$ leads to a reduced record rate and a reduced record number. We will try to estimate whether such an $\alpha\neq1$ can be detected in the data and analyze if the record statistics show signs of the AR(1) model as well.

\subsection{Records in the GARCH-model}
\label{fin:garch}

Another generalization of the simple biased random walk described above is the GARCH process. An important feature of the random walk is that its increments, the jumps $\xi_i$, are sampled from a distribution $f\left(\xi\right)$ with a fixed shape and a constant standard deviation. However, in actual stock data the standard deviation of these jumps, for instance of daily returns, can fluctuate randomly over time (see Fig. \ref{Fig:Stocks_returns_gauss_sp}). 

In 1986, Bollerslev \cite{Bollerslev1986} proposed a so called generalized autoregressive conditional heteroscedasticity (GARCH) model to describe and analyze these fluctuations in the time-dependent standard deviation (often called volatility) of stock returns. The entries $X_0,X_1,...,X_n$ of an (unbiased) GARCH($p$,$q$) process are given by $X_0=0$ and 
\begin{eqnarray}\label{Garch1}
 X_{i} = X_{i-1} + \sigma_i \xi_i,
\end{eqnarray}
where $\sigma_i$ is the standard deviation of the increment at time $i$. This standard deviation is obtained via a second, recursive equation:
\begin{eqnarray}\label{Garch2}
 \sigma_i^2 = \alpha_0 + \sum_{j=1}^p \alpha_j \left(\xi_{i-j}\sigma_{i-j}\right)^2 + \sum_{k=1}^q \beta_k \sigma_{i-k}^2
\end{eqnarray}
with a set of parameters $\left(\alpha_0,\alpha_1,...,\alpha_p,\beta_1,...,\beta_q\right)$. This more comprehensive model was specifically designed to model financial data and has become a very popular tool in financial modeling \cite{Bollerslev1992,Bera1993,Lamoureux1990,Duan1995}. 

\begin{figure*}
\includegraphics[width=0.48\textwidth]{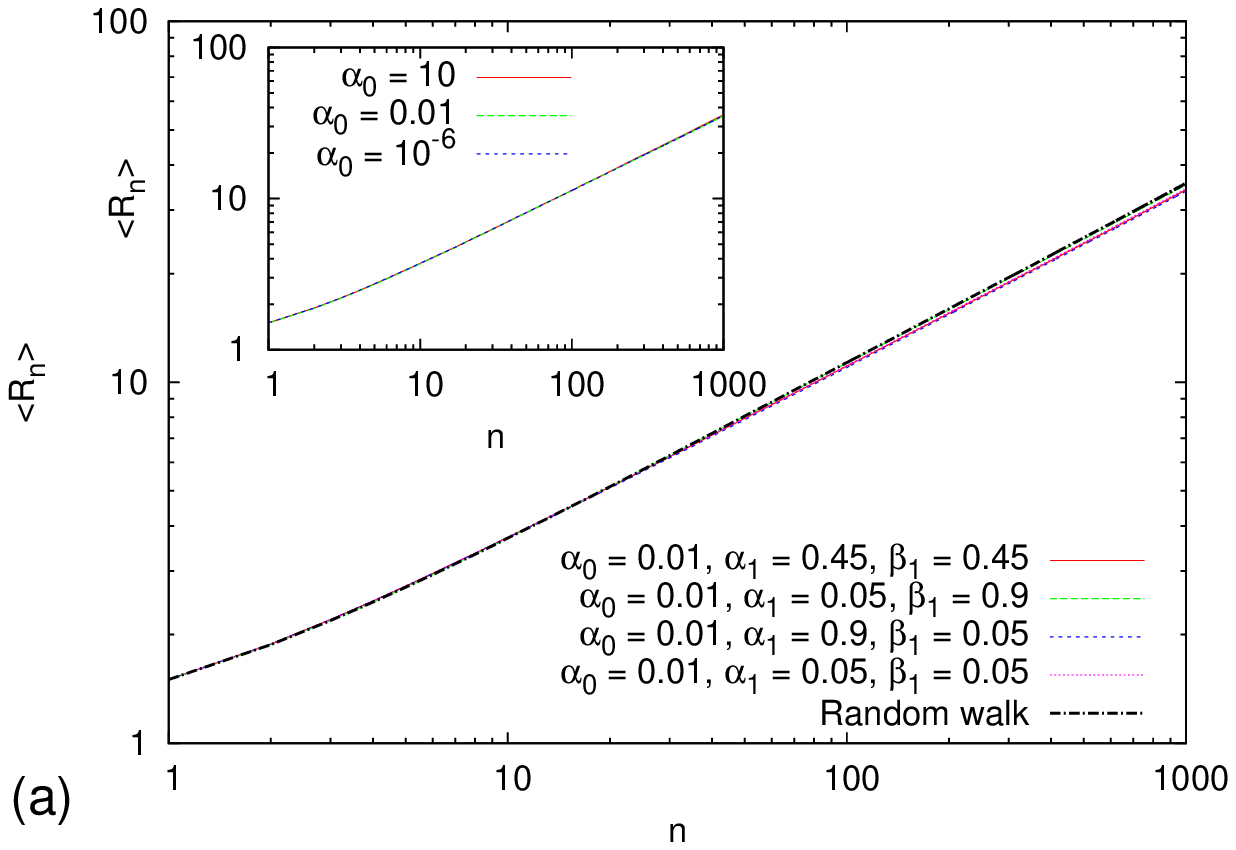}
\includegraphics[width=0.48\textwidth]{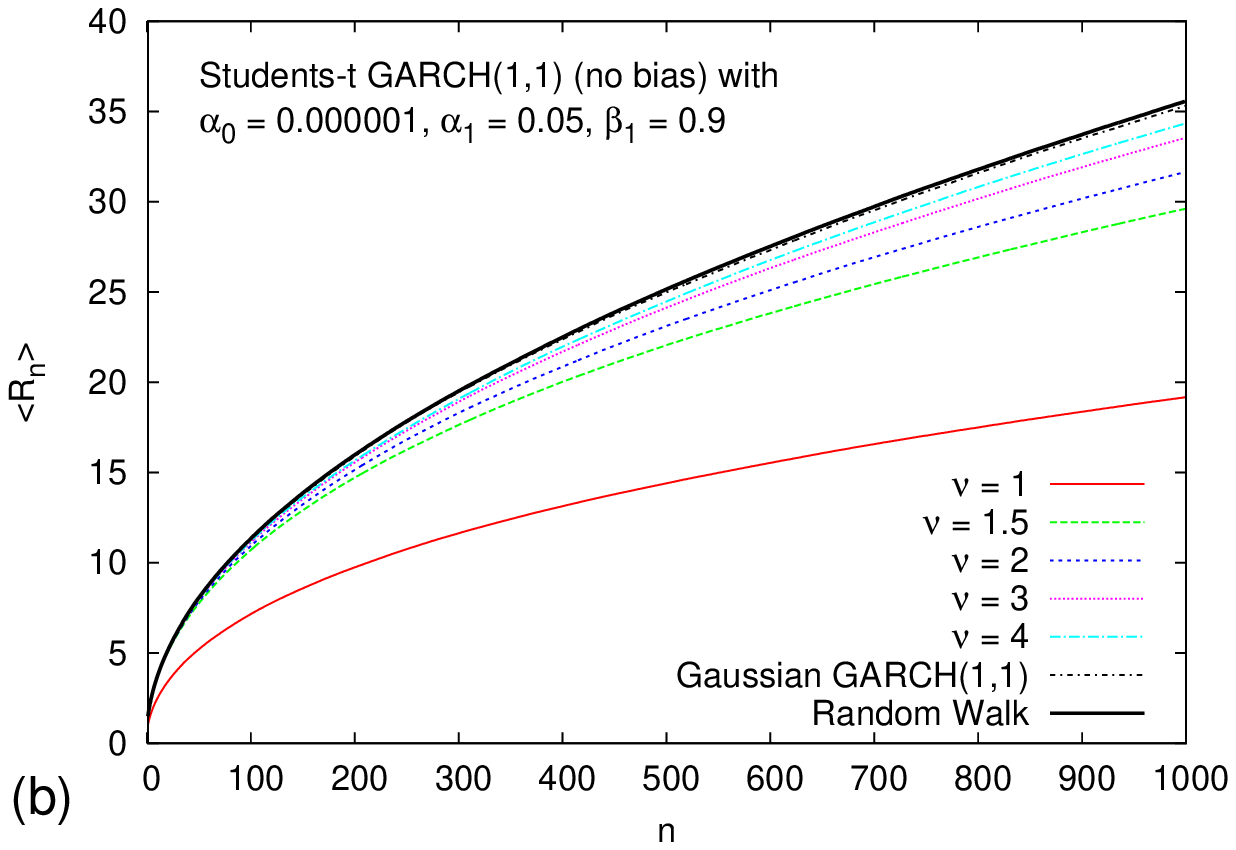}
\caption{\label{Fig:GARCH_sim_parameters} (Color online) (a) Numerical simulations the mean record number $\langle R_n\rangle$ of a Gaussian GARCH(1,1) process as described in the text (see Eqs. \ref{Garch1},\ref{Garch2},\ref{Garch3}). For each set of parameters we performed $10^5$ simulations of a GARCH process with $n=10^3$ steps. The curves in the main plot show the upper mean record number $\langle R_n \rangle$ for various choices of the two parameters $\alpha_1$ and $\beta_1$ with $\alpha_1+\beta_1<1$. The black dashed line give the random walk result ($\alpha_1=\beta_1=0$). In the inset the parameter $\alpha_0$ is varied. In all considered cases the behavior of the mean record number does not differ significantly from the random walk case. (b) Numerical simulations of the mean record number $\langle R_n\rangle$ of a GARCH(1,1) process as described in the text (see Eqs. \ref{Garch1}, \ref{Garch2}, \ref{Garch3}) with a Students-T jump distribution (see Eq. \ref{StudentsT}). For each set of parameters we performed $10^5$ simulations of a GARCH(1,1) process with $n=10^3$ steps. We computed the mean record number $\langle R_n\rangle$ for different values of the parameter $\nu=1,1.5,2,3$ and $4$ of the Students-T distribution for a fixed set of parameters $\alpha_0=0.000001,\alpha_1=0.05$ and $\beta_1=0.9$. We also gave the results for the Gaussian GARCH(1,1) (dashed line) with the same parameters $\alpha_0,\alpha_1$ and $\beta_1$ as well as the random walk result with $\alpha_1=\beta_1=0$ (thick black line). The numerical results do not depend strongly on the choice of the three parameters $\alpha_0, \alpha_1$ and $\beta_1$ as long as $\alpha_1+\beta_1<1$.}
\end{figure*}

For the sake of simplicity, we consider only the simple case of a GARCH(1,1) process with 
\begin{eqnarray}\label{Garch3}
 \sigma_i^2 = \alpha_0 + \alpha_1 \xi_{i-1}^2\sigma_{i-1}^2 + \beta_{1}\sigma_{i-1}^2. 
\end{eqnarray}
Unfortunately, in comparison to the random walk, already this simple case has three additional parameters $\alpha_0,\alpha_1$ and $\beta_1$. However, it is known that in this case, the process only has a stationary solution if $\alpha_1+\beta_1<1$ \cite{Bollerslev1986,Bollerslev1992}. For $\alpha_1+\beta_1<1$ and $i\rightarrow\infty$, the asymptotically stationary variance of the entries $X_i$ is given by \cite{Bollerslev1986,Bollerslev1992}:
\begin{eqnarray}
 \textrm{Var}\left(X_i\right) = \frac{\alpha_0}{1-\alpha_1-\beta_1}.
\end{eqnarray}
Since an analytical computation of the record statistics of the GARCH(1,1) process is (at least to our knowledge) not possible, we performed numerous numerical simulations with various sets of parameters $\alpha_0,\alpha_1$ and $\beta_1$. Interestingly, for all considered models with a Gaussian jump distribution $f\left(\xi\right)$ the record statistics of the GARCH(1,1) seems to be identical to the record statistics of the random walk discussed above. Even though we are not able to give a proof, we believe that, at least for the unbiased case, the record statistics of the  Gaussian GARCH(1,1) equals the one of the random walk. It seems that, for $n\gg1$ and independent of $\alpha_0,\alpha_1$ and $\beta_1$ with $\alpha_1+\beta_1<1$, the mean record number and the record rate of the Gaussian GARCH(1,1) process are given by
\begin{eqnarray}
  \langle R_n^{\left(\alpha_0,\alpha_1,\beta_1\right)} \rangle \approx \frac{2\sqrt{n}}{\pi} \quad \textrm{and} \quad P_n^{\left(\alpha_0,\alpha_1,\beta_1\right)} \approx \frac{1}{\sqrt{\pi n}}.
\end{eqnarray}
In the left panel of Fig.~\ref{Fig:GARCH_sim_parameters}, we show the behavior of the mean record number of a Gaussian GARCH(1,1) for several arbitrary choices of the three parameters $\alpha_0,\alpha_1$ and $\beta_1$. In all considered cases, the curves are indistinguishable from the random walk result.

Beyond this interesting observation, we found that the record statistics of the GARCH(1,1) depends crucially on the shape, and in particular the tail-behavior of the underlying distribution. With an application to stock data in mind, we decided to analyze the record statistics of the GARCH(1,1) process for the popular Students-T distribution \cite{Gosset1908,Fisher1925}, which is a well known distribution in financial modeling \cite{Baillie1990,Bollerslev1992,Brooks2008}. This distribution is of the form
\begin{eqnarray}\label{StudentsT}
f_{\nu}\left(\xi\right) = \frac{\Gamma\left[\left(\nu+1\right)/2\right]}{\sqrt{\nu\pi}\Gamma\left[\nu/2\right]} \left(1+\frac{\xi^2}{\nu}\right)^{-\left(\nu+1\right)/2}.
\end{eqnarray}
with a parameter $\nu>0$. For a finite $\nu$ the tails of this symmetric, bell-shaped distribution decay proportional to $|\xi|^{-\nu-1}$. Its Fourier transform is of the form $\tilde{f}\left(k\right) \approx 1-\left(\alpha_{\nu}|k|\right)^{-\nu}$, which, if $\nu$ is smaller than $2$, equals the $\tilde{f}\left(k\right)$ of the L\'evy flights discussed above. In this case $\nu$ equals the L\'evy index $\mu$. For $\nu\rightarrow\infty$ on the other hand, $f_{\nu}\left(\xi\right)$ approaches the standard normal distribution. 
 
Using the Students-T distribution, we analyzed the dependence of the mean record number of the GARCH(1,1) on the tail-behavior of the jump distribution. In the right panel of Fig.~\ref{Fig:GARCH_sim_parameters} we computed the (upper) mean record number $\langle R_n\rangle$ of a GARCH(1,1) process for a fixed set of parameters $\alpha_0,\alpha_1$ and $\beta_1$ with $\alpha_1+\beta_1<1$. Apparently, the mean record number of the process decreases when the tails of the jump distribution become heavier. For the very heavy-tailed Cauchy distribution (which is obtained for $\nu=1$), the mean record number after $n=1000$ time steps is only half as large as in the random walk case. Further numerical simulations showed that the dependence of the mean record number on the parameters $\alpha_0,\alpha_1$ and $\beta_1$ (with $\alpha_1+\beta_1<1$) is much less significant. The record statistics of the GARCH(1,1) process seems to be dominated by the tail-behavior of the jump distribution $f\left(\xi\right)$. Having the results of the previous section about the AR(1) process in mind, the heavy-tailedness of the jump distribution in the GARCH(1,1) models plays a similar role as the autoregressive component in the AR(1) model that was introduced with the parameter $\alpha<1$. Both generalizations of the simple random walk model decrease the mean record number $\langle R_n\rangle$.

In our discussion of stock data, we will determine the tail-behavior of the jump distribution as well as a set of parameters $\alpha_0,\alpha_1$ and $\beta_1$ from the stock data via maximum-likelihood methods and discuss the record statistics of the GARCH(1,1) model with these particular properties in more detail. 

\section{Analysis of single stocks}
\label{fin:single_stocks}

\subsection{Data introduction}
\label{fin:data_introduction}

As in \cite{Wergen2011b}, we consider a data set of daily prices of stocks contained in the Standard and Poors 500 (S\&P 500) index \cite{SPdata}. The S\&P 500 is an important stock market index and includes mostly U.S. companies, which are selected by a committee. With a market capitalization of more than $10^4$ billion USD, the index is supposed to represent all relevant branches of the U.S. industry. 

\begin{figure*}
\includegraphics[width=0.48\textwidth]{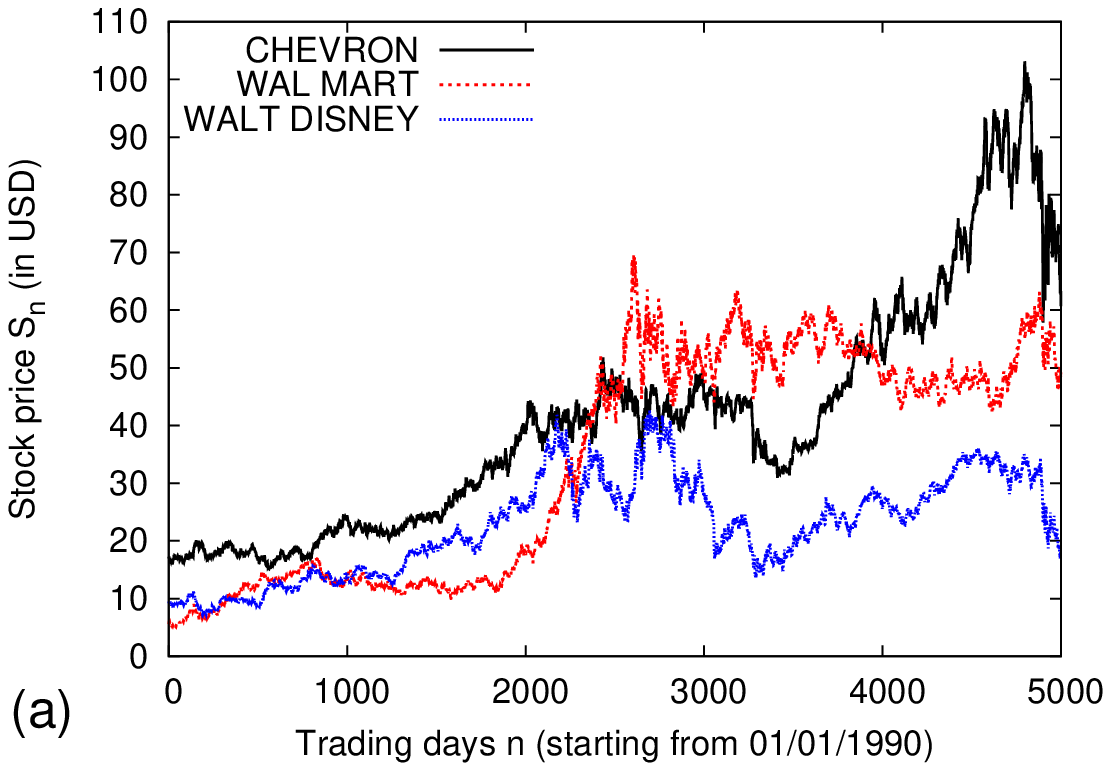}
\includegraphics[width=0.48\textwidth]{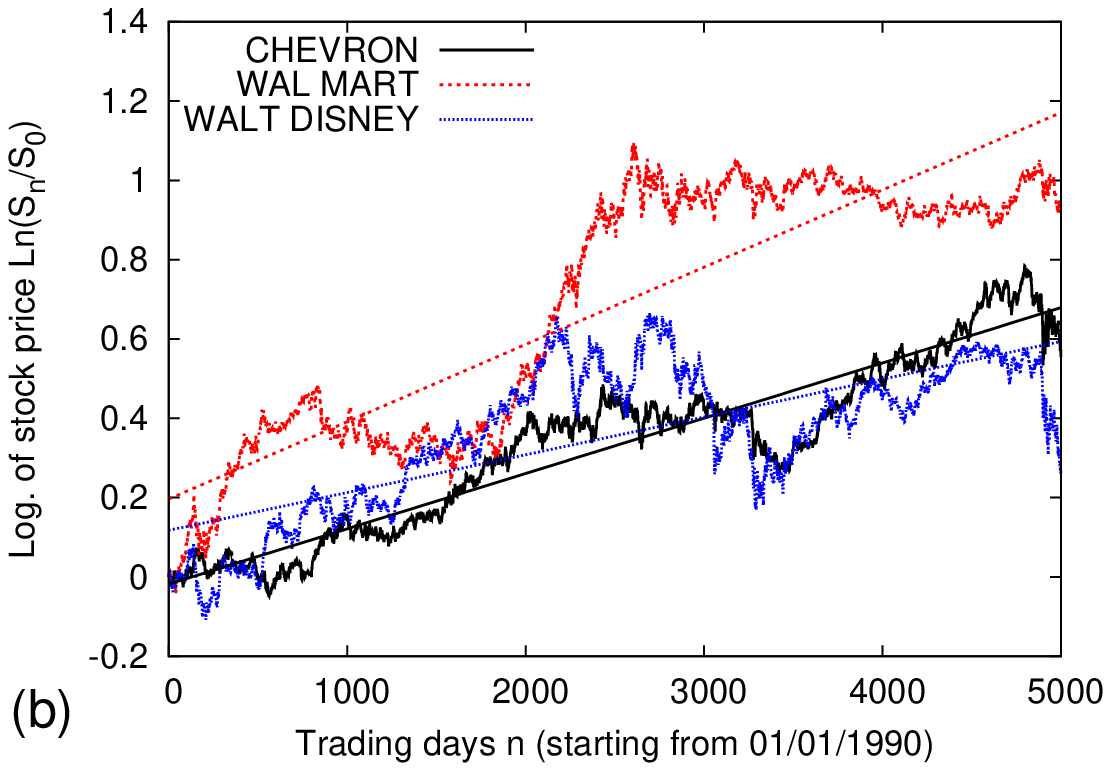}
\caption{\label{Fig:Stocks_undetr_detr} (Color online) (a) The daily stock prices $S_n$ (in Dollar) of three stocks (Chevron, Wal Mart and Disney) from the S\&P 500 index. (b) The logarithms $s_n=\ln\left(S_n/S_0\right)$ for the same three stocks along with linear regressions for these logarithmic stock prices. }
\end{figure*}

While the index is computed as a weighted average of the stock prices depending on how many shares of the stocks are publicly tradable, we will consider only the prices of the contained stocks themselves and not the score of the index. To further simplify our studies, we examine a set of daily closing prizes of $366$ stocks that remained within the S\&P 500 index for the entire time span from January 1990 to March 2009. This allows us to analyze $366$ time series with an identical length of $n=5000$ trading days.

Probably the oldest model of stock prices is the so-called Geometric Random Walk Model (GRM) \cite{MantegnaStanley,Voit}, which was introduced already more than 100 years ago by Le Bachelier \cite{LB}. In this model the logarithms $s_n:=\ln \left(S_n/S_0\right)$ of the stock prices perform a random walk with a constant bias:
\begin{eqnarray}
 s_n = s_{n-1}  + \xi_n + c.
\end{eqnarray}
This bias $c$ is supposed to represent some kind of inherent growth like a long-term interest rate or an exponentially growing amount of money in the market. Fig.~\ref{Fig:Stocks_undetr_detr} illustrates this model with three randomly selected stocks from the S\&P 500 index. It is important to notice that the occurrence of records in the stocks is not affected by the logarithm. Due to the monotony of the logarithm, we have
\begin{eqnarray}
 S_n & > & \textrm{max}\{S_0,S_1,...,S_{n-1}\} \nonumber \\ & \Leftrightarrow & s_n  >  \textrm{max}\{ 0,s_1,...,s_{n-1}\}
\end{eqnarray}
and therefore a record breaking stock prices $S_n$ is also a record in the series of the logarithms $s_n=\ln \left(S_n/S_0\right)$. On the other hand, a record of the daily returns $\Delta_n = S_n - S_{n-1}$ is not necessarily a record of the logarithmic returns $\delta_n:=s_n-s_{n-1}$. However, it is easy to show that if a logarithmic return $\delta_n$ is a record, this return is also a new record in the series of relative daily changes of the stock price $S_n/S_{n-1}$:
\begin{eqnarray}
 \frac{S_n}{S_{n-1}} & > & \textrm{max}\{0,\frac{S_1}{S_0},...,\frac{S_{n-1}}{S_{n-2}}\} \nonumber \\ & \Leftrightarrow &\quad \delta_n > \textrm{max}\{0,\delta_1,...,\delta_{n-1}\}.
\end{eqnarray}
Since these relative return records are usually more interesting in a growing stock market, we will consider them in the following. In the context of the GRM, the logarithmic daily returns $\delta_n$ should be i.i.d.~RV's sampled from a symmetric distribution plus a constant linear drift $c$ ($\delta_n \equiv \xi_n + c$).
 
For some applications, we will also detrend the logarithmic stocks, i.e.~a linear trend is subtracted from the logarithms $s_n=\ln S_n/S_0$ to obtain stock prices more comparable to symmetric random walks. Of course, in this case, the record statistics of the stocks are altered by removing the trend. A linear regression analysis of the individual stocks for the entire time-span of 5000 trading days gives an average normalized drift of $c/\sigma \approx 0.025$, where $\sigma$ is the standard deviation of the distribution of the returns $\delta_n$.

\begin{figure*}
\includegraphics[width=0.48\textwidth]{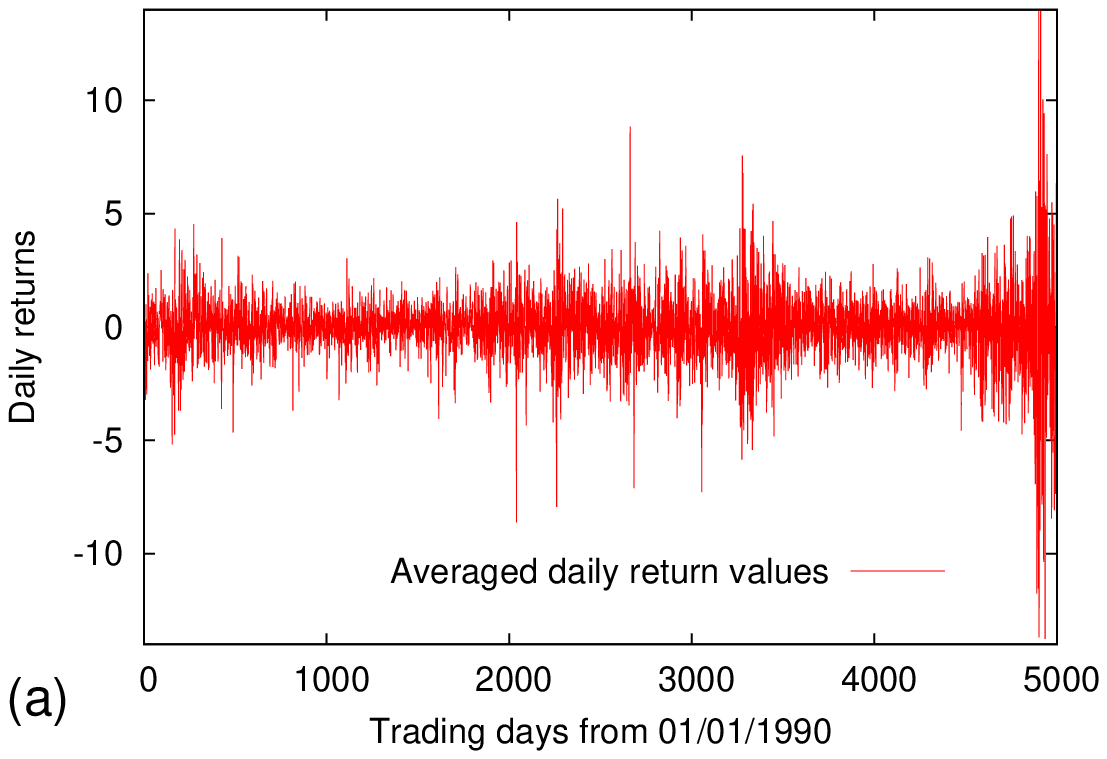}
\includegraphics[width=0.48\textwidth]{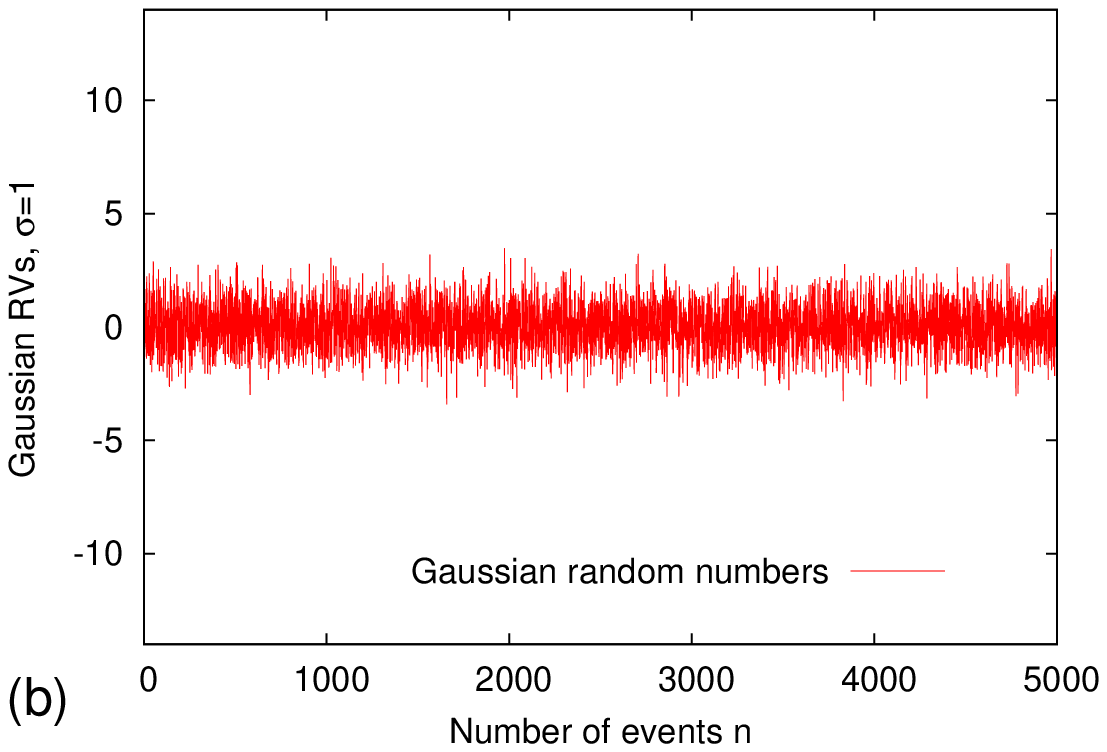}
\caption{\label{Fig:Stocks_returns_gauss_sp} (Color online) (a) The averaged and normalized daily returns $\delta_n=\ln S_n/S_{n-1}$ of the stocks in the S\&P 500 index. (b) 5000 Gaussian random numbers with standard deviation unity. }
\end{figure*}

\subsection{Jump distribution of the S\&P 500}

Today, it is well known that this simple geometric random walk model does not represent a complete and accurate model of the stock markets. In fact, it is particularly useless in times of high market activity and during crashes since it assumes that the daily changes in a stock prize are random and uncorrelated. Actual stocks are of course correlated with other stocks and, in addition, usually non-Markovian \cite{MantegnaStanley,Voit}. 

In Fig.~\ref{Fig:Stocks_returns_gauss_sp}, we compare the 5000 averaged and normalized logarithmic daily returns $\delta_i / \sigma$ (where $\sigma$ is the standard deviation of the return distribution) of the S\&P 500 with 5000 computer-generated Gaussian random numbers with standard deviation unity. While the pattern of the Gaussian RV's is homogeneous, the amplitudes of the stock returns fluctuate over time. Despite these findings, many statistical properties of stock prices can be modeled and understood using the GRM and, as it was already shown in previous studies, it is also, to some degree, useful to model the record statistics of stocks in the S\&P 500. 

\begin{figure}
\includegraphics[width=0.48\textwidth]{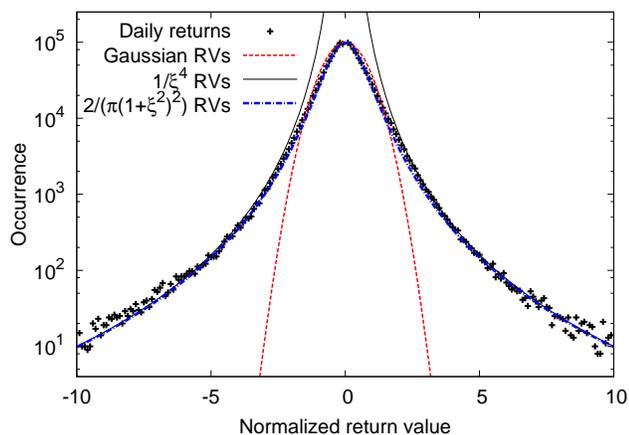}
\caption{\label{Fig:detr_norm_return_distr_log} (Color online) The distribution of the normalized and detrended logarithmic daily returns of the stocks in the S\&P 500. We considered the daily returns $\tilde{\delta_i} = \left(\delta_i-c\right)/\sigma$ of the logarithmically detrended stocks, where the drift $c$ and the standard deviation $\sigma$ were obtained from the stock data. These normalized returns have standard deviation unity. The plot was obtained by binning the values into bins of size $0.01$. For comparison, we also plotted a manually fitted Gaussian (red dashed line) with standard deviation $\sigma=1$ and a symmetric Pareto distribution with $f\left(\xi\right) = 1/\xi^4$ (black line). Apparently, the tails of the detrended data are not Gaussian and instead well described by the $1/\xi^{4}$-power-law. Furthermore we plotted the normalized Students-T distribution (blue dashed line) with $\nu=3$ (see Eq. \ref{fin:studentsTnorm}), which closely resembles the distribution of the stock returns.}
\end{figure}

Since, in the context of the record statistics of biased random walks, the shape of the jump distribution is of importance, we measured the probability density of the logarithmic daily returns $\delta_n$. In Fig.~\ref{Fig:detr_norm_return_distr_log}, the probability density of the these returns, after linearly detrending the logarithms of the stock prizes, is plotted. Apparently, while the daily returns close to zero are approximately normally distributed, their tails are highly non-Gaussian and decay like a power-law with $1/\xi^{4}$. In spite of that, this return distribution still has a finite second moment. The corresponding L\'evy index to the discovered power-law is $\mu=3$ and is much larger than the critical value of $\mu=2$ (see section \ref{fin:random_walks}). A random walk with such a jump distribution is not a L\'evy flight. 

In the context of our discussion of the GARCH(1,1) model with a Students-T jump distribution, we can now give the shape of the Students distribution suitable for modeling the daily returns of our stock data. Because of Fig.~\ref{Fig:detr_norm_return_distr_log}, the returns should be well described by a Students-T distribution with parameter $\nu=3$, which leads to $f_3\left(\xi\right) = 6\sqrt{3}\left(\pi\left(3+\xi^2\right)^2\right)$. Normalized to standard deviation unity this distribution assumes the following form:
\begin{eqnarray}\label{fin:studentsTnorm}
 f\left(\xi\right) = \frac{2}{\pi\left(1+\xi^2\right)^2}.
\end{eqnarray}
This distribution is also compared to the normalized logarithmic returns of the stock data from the S\&P 500 in Fig.~\ref{Fig:detr_norm_return_distr_log}. While the simple power-law with $f\left(\xi\right)=1/\xi^{4}$ is only accurate in the tails of the return distribution, the distribution given in Eq.~\ref{fin:studentsTnorm} seems to describe the entire distribution to a good accuracy.

Knowing that the return distributions of stock prizes on shorter time-scales, such as minutes or seconds, are much broader than the distributions of daily returns \cite{MantegnaStanley,Voit}, the results of Majumdar et al.~\cite{Majumdar2012} for L\'evy-indices $\mu<2$ might still be useful in future studies of the record statistics of stock prices with a higher temporal resolution.

\subsection{Record statistics of the increments}

\begin{figure*}
\includegraphics[width=0.48\textwidth]{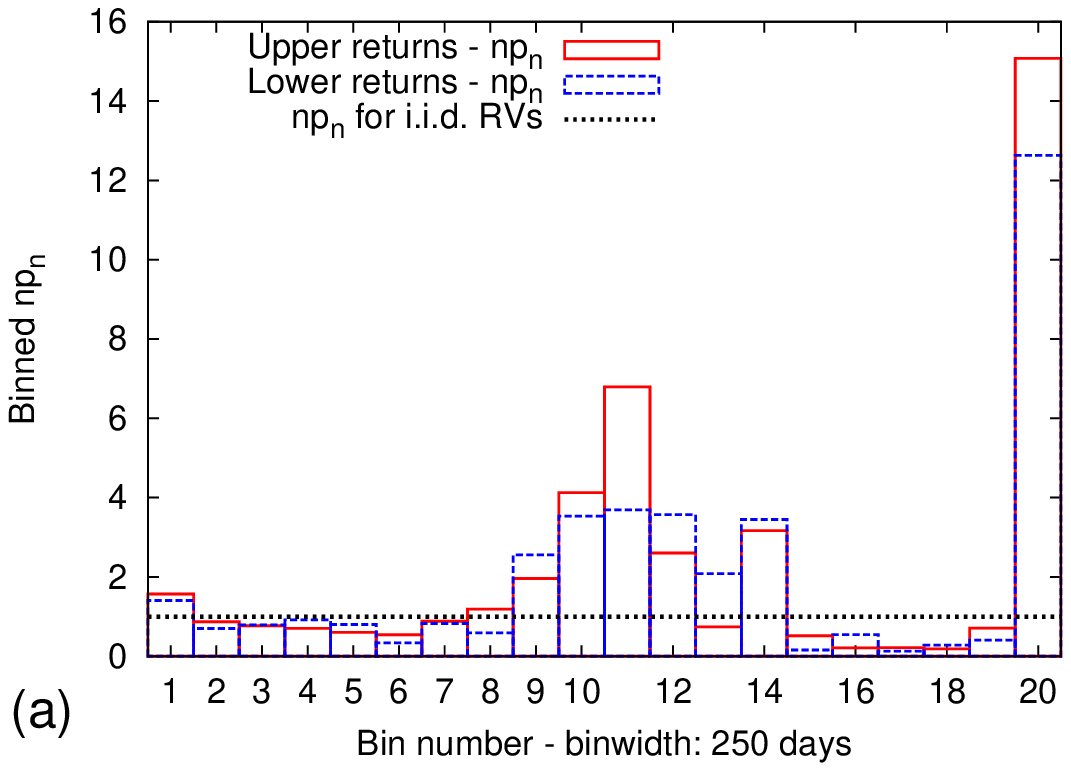}
\includegraphics[width=0.48\textwidth]{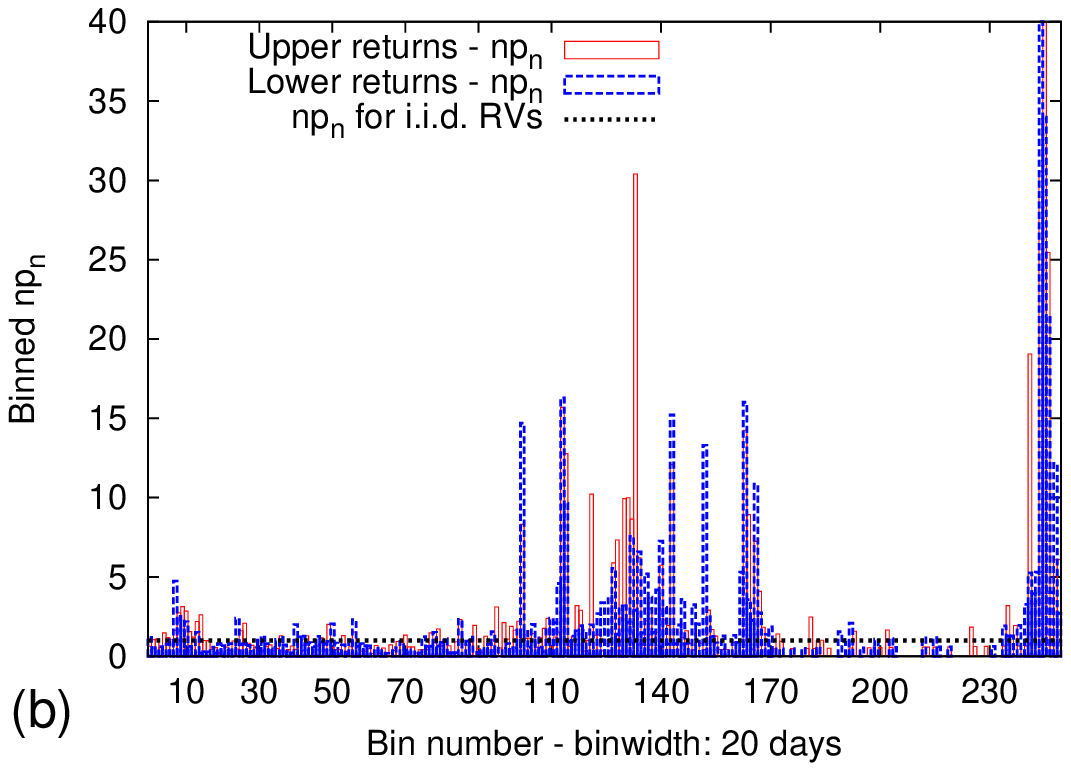}
\caption{\label{Fig:return_record_rate_binned} (Color online) (a) Binned normalized record rate of the daily returns $\delta_n = \ln S_n/S_{n-1}$ in the S\&P 500. We computed the normalized record rate $np_n$ in bins of 250 trading days length. For each bin, we counted the number of upper and lower records of $\delta_n$ and multiplied it with the number of the bin. As discussed in the main text, the upper and lower record rates are highly correlated. (b) The same analysis but for 250 bins of 20 trading days length. Note the extremely high number of new upper and lower return records during the financial crisis in 2009. In both considered cases the return record rate fluctuates strongly and deviates significantly from the i.i.d. results ($np_n=1$). The last peak in the lower figure exceeds $np_n=95$ for the upper return records.}
\end{figure*}

We begin our study of the record statistics of stocks in the S\&P 500 with an analysis of record-breaking daily returns. On the basis of the simple Geometric Random Walk Model (GRM), the returns $\delta_n=\ln S_n/S_{n-1}$ should have the same record statistics as i.i.d.~RV's. Therefore, the record process of these increments gives an opportunity to check if the returns of the stock prices and i.i.d.~RV's are comparable. If we find that the returns have the same record statistics as i.i.d.~RV's this would indicate that they are also more or less uncorrelated and have no systematic time-dependence.

We analyzed the records in the logarithmic daily returns $\delta_n$ and computed the mean record number $\langle r_n \rangle$ in these time series. At the end of the entire period of $5000$ trading days, we found an average number of $11.09$ upper and $11.12$ lower records. For i.i.d.~RV's we would have expected only $\langle r^{\left(\textrm{iid}\right)}_n\rangle \approx 9.094 \pm 0.389$ records, so, at least on such a long time-scale, the record statistics seems to differ from the one of i.i.d.~RV's. However, the actual error margins might be much larger since we assumed $366$ independent stocks to compute them. It turns out that, for this long time-span of $5000$ trading days, the record statistics of returns is dominated by fluctuations and a significant amount of the records that contribute to the mean record number are set on a very small number of trading days.

In Fig.~\ref{Fig:return_record_rate_binned}, we illustrate how the record rate of the returns varies over time. The figure shows a normalized version of the return record rate. We computed the rate $p_n$ of new return records of $\delta_n$ for each trading day and multiplied this rate with the number of the trading day $n$. For i.i.d.~RV's one expects $n p_n^{\left(\textrm{iid}\right)}=1$ for an arbitrary value of $n$ and, therefore, we call $np_n$ the normalized record rate. In Fig.~\ref{Fig:return_record_rate_binned}, the distribution of $np_n$ for the returns $\delta_n$ of the 366 stocks is binned in bins of 250 (left panel) and 20 (right panel) trading days length. The left panel of Fig.~\ref{Fig:return_record_rate_binned} shows roughly the annual distribution of the normalized return record rate $np_n$. Both the upper and the lower record rate of $\delta_n$ were very high in the years between 1998 and 2003, as well as in 2008 and 2009. In some years the record rate was more than 5 times as high as expected on the basis of i.i.d.~returns. Interestingly, the plot shows that the upper and lower record rates are highly correlated. With one exception, these rates never differ by more than $20\%$, a result that is not found for i.i.d.~RV's, where upper and lower records are uncorrelated.

In the right panel of Fig.~\ref{Fig:return_record_rate_binned}, we plotted $np_n$ for the daily logarithmic returns $\delta_n$ with a smaller bin length of 20 trading days, which is roughly one calendar month. This figure shows how extremely the record rate of $\delta_n$ fluctuates over time. A significant amount of the return records was set in only a few trading days. 

\begin{table}
\begin{center}
\begin{tabular}{|c||c||c|c|c|c|c|}
\hline
\quad\quad & Up. Recs. & Low. Recs. & \quad \quad & Up. Recs. & Low. Recs \\
\hline\hline
2040 & 0 & \textbf{156} & 4752 & 0 & 15 \\
2041 & \textbf{54} & 4 & 4880 & 0 & 13 \\
2260 & 0 & \textbf{85} & 4883 & 22 & 0 \\
2266 & 37 & 0 & 4884 & 20 & 0 \\
2293 & 28 & 0 & 4890 & 0 & \textbf{50} \\
2663 & 49 & 0 & 4891 & 14 & 0 \\
2684 & 0 & 27 & 4898 & 0 & 16 \\
2873 & 0 & 25 & 4900 & \textbf{83} & 0 \\
3055 & 25 & 0 & 4902 & 0 & 26 \\
3277 & 0 & 15 & 4911 & 25 & 0 \\
  \hline
\end{tabular}
\end{center}
 \caption{ \label{Tab:top_return_days} The 20 trading days with the highest normalized (upper or lower) return record rate $np_n$ of the daily returns $\delta_n$ together with the corresponding number of upper and lower records. The top 20 days are ordered chronologically. Days with 40 upper or lower return records were additionally highlighted. Note that 9 of these 20 trading days with the highest relative daily changes in the stock prices $S_n$ occurred during only 31 trading days (in the fall of 2008). Another 7 occurred in only 3 years following trading day $n=2040$ (1998-2000).}
\end{table}

In Tab.~\ref{Tab:top_return_days}, we list the 20 trading days with the highest normalized (positive or negative) record rate $np_n$ along with the total number of (upper and lower) return records of $\delta_n$ for these trading days. The table shows how strongly the record statistics of the returns is affected by a few periods of very high market activity. 9 of these 20 \emph{record}-days fall into a short time-period of only 31 trading days in the fall of 2008, in which the recent financial crisis had its climax. 

\begin{figure*}
\includegraphics[width=0.48\textwidth]{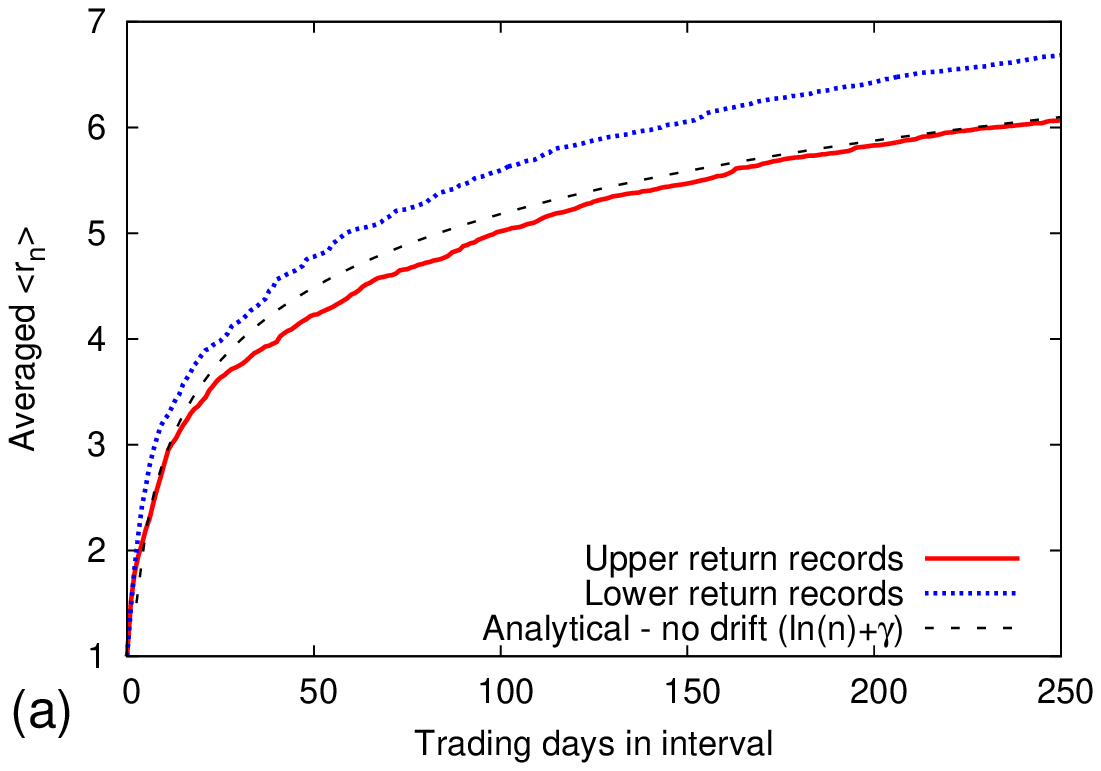}
\includegraphics[width=0.48\textwidth]{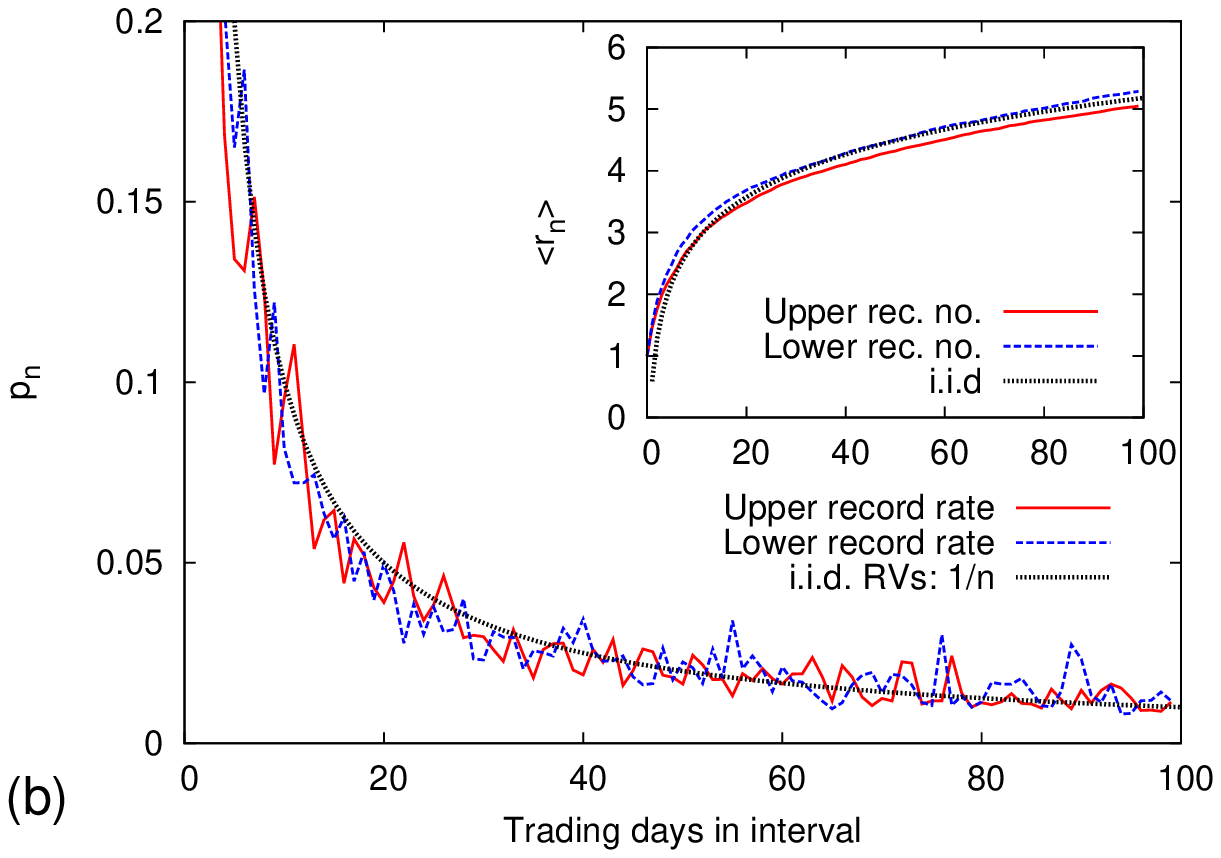}
\caption{\label{Fig:return_records_up_low_250} (Color online) (a) Averaged mean upper (red) and lower (blue) record number of the daily returns $\delta_n = \ln S_n/S_{n-1}$ in the S\&P 500. The data set was split up into 20 consecutive intervals of each 250 trading days. We computed the evolution of the mean record number of the 366 stocks separately in each of the intervals before we averaged over all intervals. The black dashed line gives the analytical result for the mean record number of i.i.d.~RV's with $\langle r_n\rangle \approx \ln n + 0.577...$. (b) Averaged upper (red) and lower (blue) record rate of the daily returns $\ln S_n/S_{n-1}$ in the S\&P 500. Here, the data was split up into 50 consecutive intervals of each 100 trading days. Again, we first computed the record rates for all 366 stocks in the individual intervals before we averaged over all intervals. The black dashed line gives the theoretical i.i.d.~result with $P_n^{\left(\textrm{iid}\right)}=1/n$. The inset shows the same analysis for the averaged mean record number.}
\end{figure*}

The table also explains the high correlations between upper and lower records, which was found for the binned data of the normalized record rate $np_n$. While there are no trading days with both, a very high number of upper and a very high number of lower records, days with many lower records are often quickly succeeded by days with a large number of upper records and vice versa. In fact, of all upper return records in the entire data set, $9.77\%$ are followed by a lower record in the next trading day. Similarly $9.33\%$ of the lower records are immediately followed by an upper record. If one considers the five days following a record, these values double. $18.00\%$ ($18.54\%$) of all upper (lower) records entail a lower (an upper) record within the next five trading days. For i.i.d.~RV's, it is easy to compute that the corresponding rates should be below $0.5\%$ on the full interval of $n=5000$ trading days.

Tab.~\ref{Tab:top_return_days} and Fig.~\ref{Fig:return_record_rate_binned} indicate that the record statistics of the returns $\delta_n$ might differ between times of relatively calm market activity and those of financial crisis. To reduce the effect of these events and to get more reliable statistics, we also analyzed the data set in shorter intervals. 

Therefore, we subdivided the 5000 trading days into 20 consecutive intervals of each 250 trading days and computed the mean record numbers of the $\delta_n$'s separately in each of these intervals before we averaged over the results. The left panel of Fig.~\ref{Fig:return_records_up_low_250} shows the progressions of the averaged upper and lower mean record number for these intervals and compares them with the i.i.d.-curve given by $\langle r^{\left(\textrm{iid}\right)}_n \rangle = \ln n + \gamma$. Apparently, the mean upper record number agrees quite well with the i.i.d.~result. The number of lower records is slightly increased.

We also subdivided the data in shorter intervals of each 100 trading days. For these intervals we have enough individual time series to plot the record rate $p_n$ of the daily returns $\delta_n$. The right panel of Fig.~\ref{Fig:return_records_up_low_250} compares the upper and lower record rates in these intervals with the i.i.d.~prediction of $p_n^{\left(\textrm{iid}\right)}=1/n$. Here, the curves for the stock returns are in good agreement with the i.i.d.~behavior. Furthermore, also the behavior of the mean record number $\langle r_n\rangle$ (inset in Fig.~\ref{Fig:return_records_up_low_250}b) agrees almost perfectly with the i.i.d.~result.

In summary, over longer time-spans of several years the record statistics of the logarithmic daily returns $\delta_n = \ln S_n/S_{n-1}$ of the stocks in the S\&P 500 differs significantly from the behavior of i.i.d.~RV's. Nevertheless, this effect seems to be caused by a few short periods of high market activity, in which large numbers of stocks collectively set new return records. On shorter time frames the daily returns are more similar to i.i.d.~RV's, here both the record rate $p_n$ and the mean record number $\langle r_n\rangle$ of the $\delta_n$'s are modeled accurately by the i.i.d.~results. In the context of the GRM, we have therefore reason to believe that, on these shorter intervals, the record statistics of the logarithmic stock prices $s_n = \ln S_n/S_0$ is accurately modeled by a biased random walk. 

\subsection{Record statistics of the stocks}
\subsubsection{Occurrence of records}

\begin{figure}[t]
\includegraphics[width=0.48\textwidth]{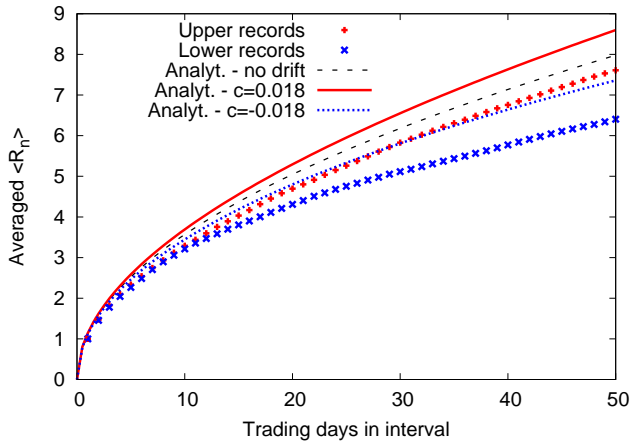}
\caption{\label{Fig:records_up_low_50} (Color online) Averaged upper (red) and lower (blue) mean record number of the daily stock prices $S_n$ in the S\&P 500. The data set was split up into 100 consecutive intervals of each 50 trading days. The records in the stock prices are compared to our analytical result for the biased random walk with a Gaussian jump distribution with standard deviation $\sigma=1$ and a drift $c=0.018$ (see Eq.~\ref{fin:Pred_finite_n}). The drift was obtained by a linear regression analysis of the logarithmic stock prices $s_n = \ln S_n/S_{n-1}$. Both the mean record numbers and the values for the drift were first computed separately for each interval and then averaged over all intervals.}
\end{figure}

With these findings for the daily returns, we can now discuss the record statistics of the stock prices $S_n$ themselves. In \cite{Wergen2011b}, we already presented an analysis of the mean record number of the undetrended stock prices in the S\&P 500 for the full time-span of $n=5000$ trading days and also an analysis of shorter intervals of length $n=100$. While the Geometric Random Walk Model (GRM) yielded an accurate description for the long interval, we found a significant reduction of the number of lower records in the detrended data for shorter intervals. 

In the context of the findings for the returns discussed above, the good agreement for the full time-span could be a coincidence. However, especially because of Fig. \ref{Fig:return_records_up_low_250}, the record statistics of the stocks on the shorter intervals should resemble the GRM. The asymmetry between upper and lower records on the interval of 100 trading days that was discovered by Wergen et al.~\cite{Wergen2011b} is not explained by this model. 

\begin{figure*}[t]
\includegraphics[width=0.48\textwidth]{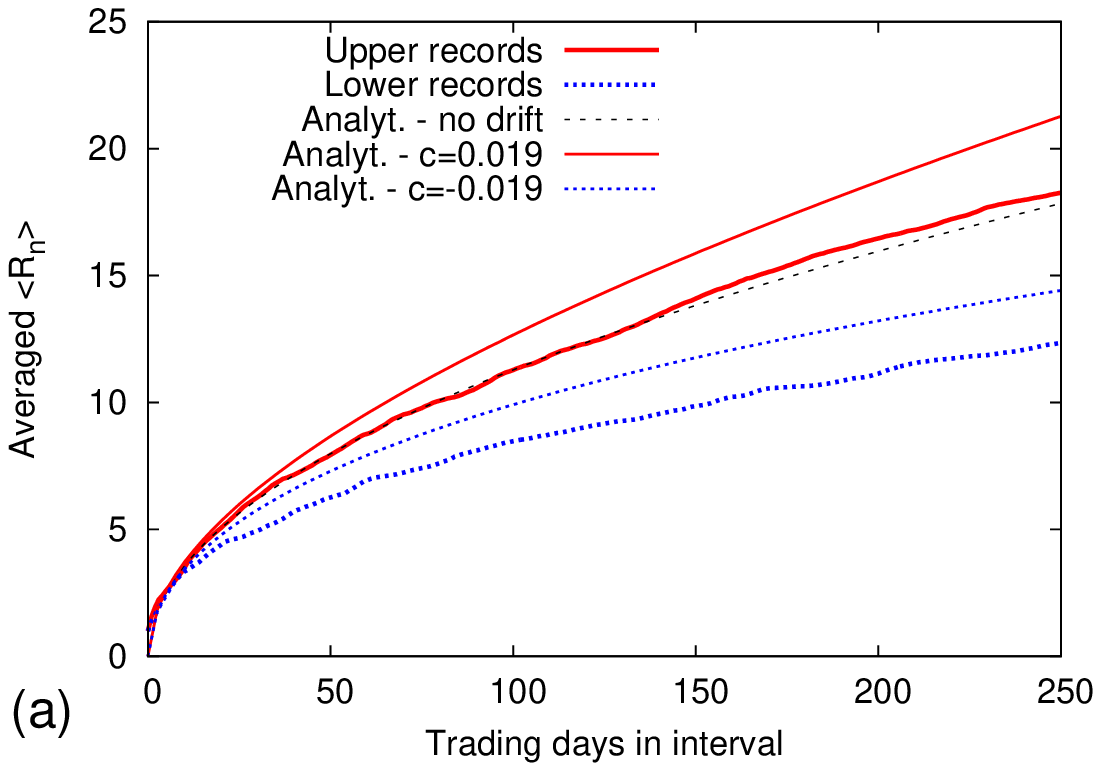}
\includegraphics[width=0.48\textwidth]{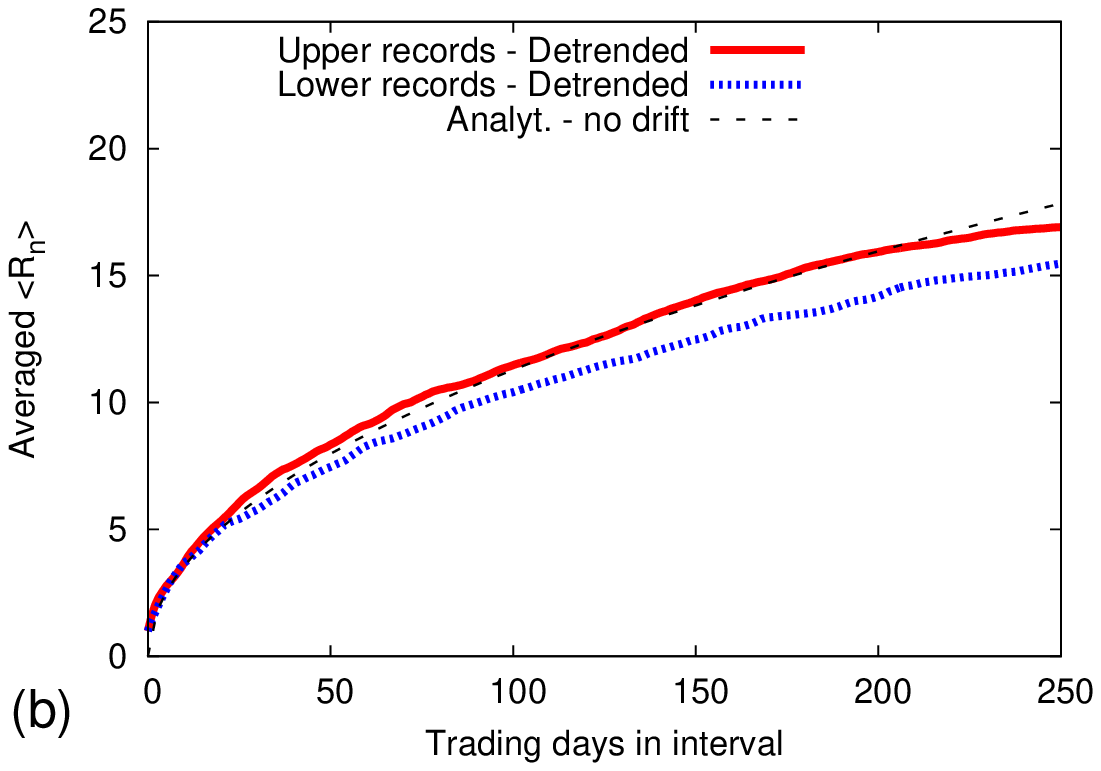}
\caption{\label{Fig:records_up_low_250} (Color online) (a) Averaged upper (red) and lower (blue) mean record number of the daily stock prices $S_n$ in the S\&P 500. The data set was split up into 20 consecutive intervals of each 250 trading days. The records in the stock prices are compared to our analytical result for the biased random walk with a Gaussian jump distribution with standard deviation $\sigma=1$ and a drift $c=0.019$ (see Eq.~\ref{fin:Pred_finite_n}). The drift was obtained by a linear regression analysis of the logarithmic stock prices $s_n = \ln S_n/S_{n-1}$. Both the mean record numbers and the values for the drift were first computed separately for each interval and then averaged over all intervals. (b) The same analysis but for detrended daily stock prices. We subtracted a linear trend from all individual stocks in the individual intervals before we computed and averaged the record numbers. The detrended data is compared to the analytical prediction for the unbiased random walk (black dashed line). }
\end{figure*}

A similar asymmetry is also found if we subdivide the data into intervals of different lengths between $20$ and $1000$ trading days. In Fig.~\ref{Fig:records_up_low_50} we splitted the series of stock prices into 100 consecutive intervals of each 50 trading days. The figure shows the averaged mean record number of the stock prices. Similarly, in Fig.~\ref{Fig:records_up_low_250} (left panel), we consider the data set subdivided into intervals of each 250 trading days, which we also considered for the daily returns. In both figures (Fig.~\ref{Fig:records_up_low_50} and Fig.~\ref{Fig:records_up_low_250}a), we compare the mean record number of the undetrended data with the analytical predictions from the GRM of a biased random walk. The analytical results were computed for a Gaussian jump distribution with standard deviation unity and drift rates of $c=0.018$ (Fig.~\ref{Fig:records_up_low_50}) or $c=0.019$ (Fig.~\ref{Fig:records_up_low_250}a) respectively. We computed the average drift $c$ from the data and plotted the curve described by Eq.~\ref{fin:Pred_finite_n}. The GRM results predict the difference between the number of upper and lower records correctly, but the observational curves lie distinctly below the theoretical ones. Apparently, both the upper and the lower record numbers in the stocks are systematically decreased in comparison to the GRM.  

The right panel in Fig.~\ref{Fig:records_up_low_250} shows the averaged upper and lower mean record number in the detrended data. The stocks were detrended separately in each interval before we analyzed and averaged the records. The records in the stocks are compared to the analytical result for the symmetric random walk with $\langle R_n\rangle \approx \sqrt{4n/\pi}$. The upper record numbers are in good agreement with the (unbiased) random walk result, but, here, in contrast to our analysis of the daily returns (see Fig.~\ref{Fig:return_records_up_low_250}), the lower record number is decreased.

\begin{figure*}
\includegraphics[width=0.48\textwidth]{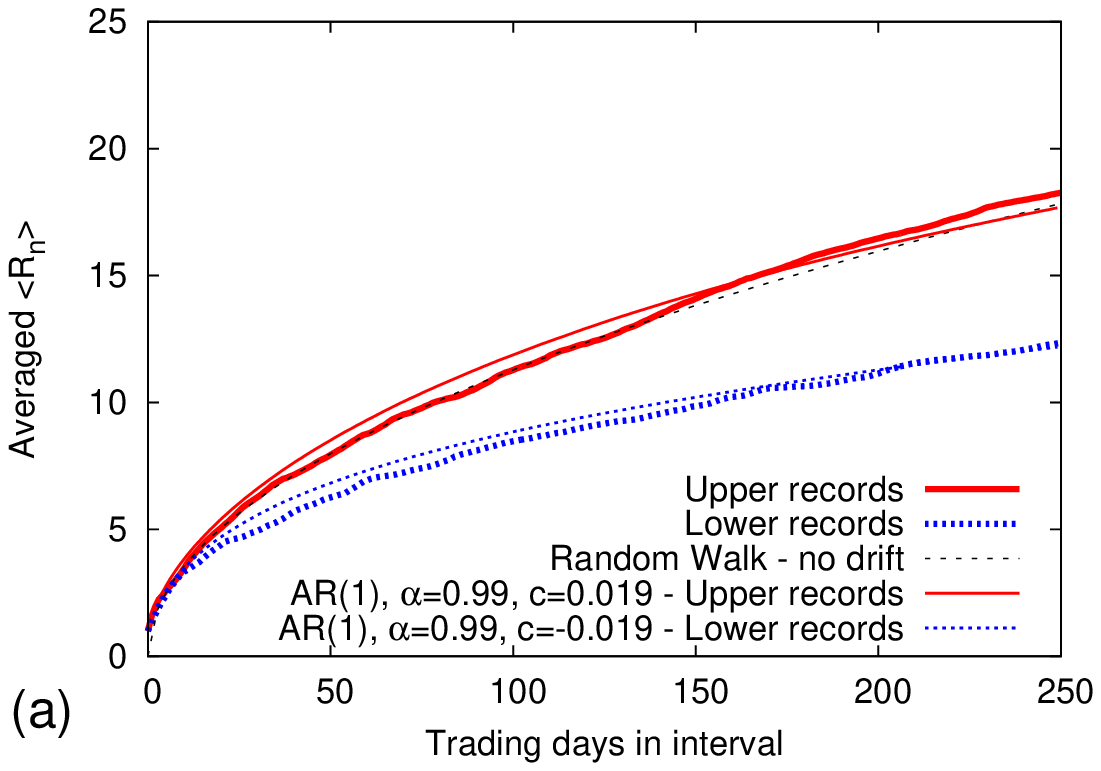}
\includegraphics[width=0.48\textwidth]{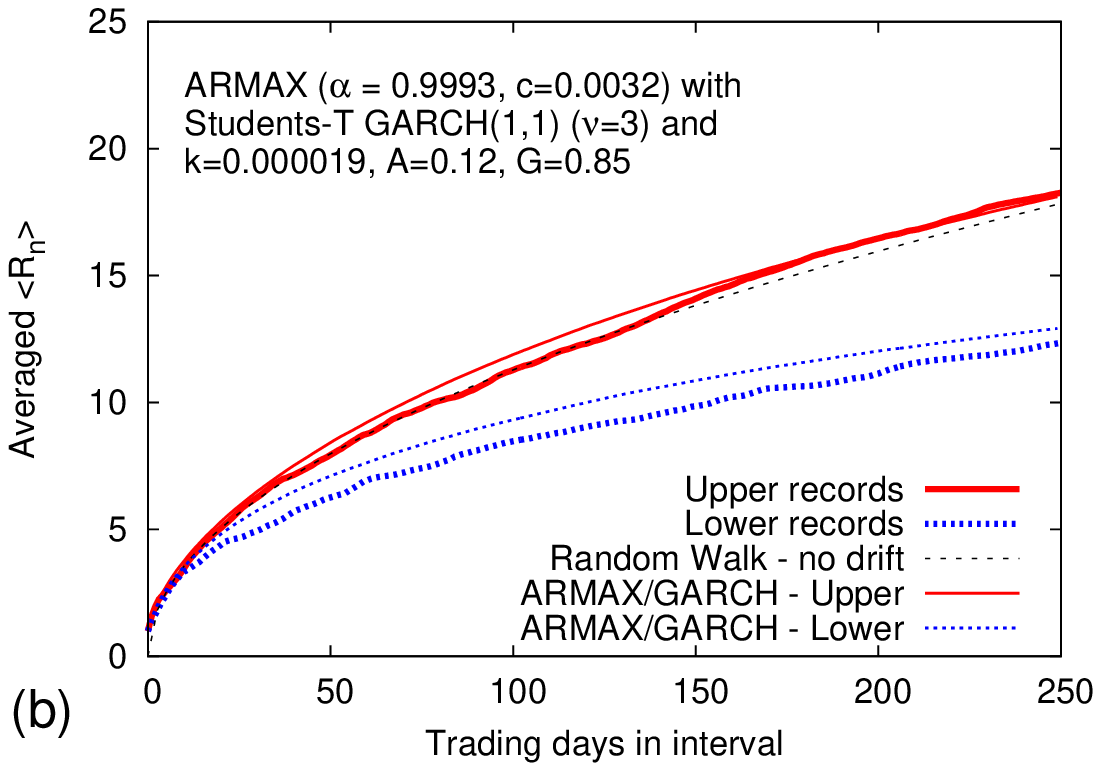}
\caption{\label{Fig:ar1_250} (Color online) (a) Averaged upper (red) and lower (blue) mean record number of the daily stock prices in the S\&P 500. The data set was split up into 20 consecutive intervals of each 250 trading days. We compare the observational data with numerical simulations for the AR(1) process with a manually fitted value for $\alpha=0.99$ and a bias of $c=0.019$ (dotted lines). (b) Here, we compare the same observational data with numerical simulations for an autoregressive model with GARCH(1,1) returns (thin red and blue dotted lines). The parameters of the model were estimated by a maximum-likelihood method, there was no manual fitting involved. The model predictions agree relatively well with the observational data.}
\end{figure*}

Since, as demonstrated in the left panel of Fig.~\ref{Fig:records_up_low_250}, the GRM of biased random walks fails partly in modeling the mean record number of daily stocks, we compared the data for the intervals of $250$ trading days to the more sophisticated AR(1) process introduced in section \ref{fin:autoregressive}. When the parameter $\alpha$ of the AR(1) process is manually set to $\alpha=0.99$, this process seems to describe the mean record numbers of the stocks more accurately. As illustrated in the left panel of Fig.~\ref{Fig:ar1_250}, both the total upper and lower record numbers and the differences between the two are modeled correctly. However, manual fitting of the parameter $\alpha$ is not entirely satisfying. A maximum-likelihood estimation, which can be performed with common data analysis software such as Matlab, yields a value of $\alpha\approx0.9993$ much closer to one. With this value of $\alpha$, the AR(1) model would not describe the record statistics of the stocks accurately. 

In a second step, we compared the mean record number of the stocks with the more complicated GARCH(1,1) model discussed in section \ref{fin:garch}. We simulated several different GARCH(1,1) models including a Gaussian GARCH(1,1) and one with a Students-T return distribution. The parameters $\alpha_0,\alpha_1$ and $\beta_1$ were obtain by a maximum-likelihood estimation. Interestingly, neither the simple Gaussian GARCH-model nor a Students-T GARCH with tail-parameter $\nu=3$ is capable of describing the behavior of the mean record number of the stock data accurately. For a Gaussian GARCH(1,1), the curves for the upper and lower mean record number were almost identical with the results we obtained for the GRM (see Fig. \ref{Fig:records_up_low_250}a). Similarly, the effect of the more heavy-tailed GARCH(1,1) with a Students-T return distribution on the difference between the upper and lower mean record number was to weak too explain the observed asymmetry.

Apparently, neither the AR(1) nor the complicated GARCH(1,1) model alone are suitable of explaining the behavior of the mean record number of the stocks in the S\&P 500. Nevertheless, since both models seem to introduce an asymmetry between the upper and lower record numbers, we decided to compare the record statistics of the stocks with a combined model with both an autoregressive and a GARCH-like component. Our maximum-likelihood estimation suggests that both components are actually present in the data. With Matlab we estimated the parameters of an autoregressive GARCH-model with entries of the following shape:
\begin{eqnarray*}
 X_i = \alpha X_{i-1} + \sigma_i \xi_i + c
\end{eqnarray*}
with Students-T increments (with $\nu=3$) and a time-dependent standard deviation of the increments given by
\begin{eqnarray*}
 \sigma_i^2 = \alpha_0 + \alpha_1\xi_{i-1}^2\sigma_{i-1}^2 + \beta_1\sigma_{i-1}^2.
\end{eqnarray*}
Using Matlab we found the following values for the five free parameters: $c=0.0032$, $\alpha=0.9993$, $\alpha_0 = 0.000019$, $\alpha_1 = 0.12$, $\beta_1 = 0.85$. It turns out that such a model predicts the behavior of the upper and lower mean record number of the stock data quite accurately. In Fig. \ref{Fig:ar1_250} (right panel), we compare numerical simulations of this autoregressive GARCH(1,1) model with a Students-T return distribution with the record statistics of the stock data. Since there is no manual fitting involved, we can assume that this agreement is more meaningful than the one obtained with manually fitting $\alpha$ in the case of the AR(1) model. A possible implication is that the combination of both, the autoregressive behavior of the stocks and the fluctuations of the standard deviation of the increments due to the GARCH model with a heavy-tailed jump distribution, lead to the observed asymmetry between the number of upper and lower records. With the parameters estimated via maximum-likelihood methods both effects alone are too weak, but when we combine them, we can describe the behavior of the mean record number of the stocks to a good accuracy.
 
\subsubsection{Correlations between stock records}

Motivated by these observations, we also considered the correlations between the record events in the individual stocks. In a simple symmetric random walk the probability $P_{n+1|n}$ of a record in the $\left(n+1\right)$th entry given that the $n$th entry was already a record is simply $P_{n+1|n} = \textrm{Prob}\left[\xi_{n+1}>0\right] = 1/2$. This is just the probability that the walker makes a positive jump. Accordingly, for a small $c\ll\sigma$, the probability for a second upper record in step $n+1$ directly following a record in step $n$ of a biased random walk is given by 
\begin{eqnarray}
 P_{n+1|n} & := & \textrm{Prob}\left[X_{n+1}\textrm{ rec.}|X_{n}\textrm{ rec.}\right] \nonumber \\ & = & \int_{-\infty}^c \mathrm{d}\xi\; f\left(\xi\right) \approx \frac{1}{2} + f\left(0\right)c > \frac{1}{2}.
\end{eqnarray}
Here, we assumed that the symmetric jump distribution $f\left(\xi\right)$ is sufficiently smooth around zero.
We compared this prediction with the daily stock prices in the S\&P 500 and found that the probability $P_{n+1|n}$ in these series is always significantly smaller than predicted by the GRM. On average, the correlations both for upper and lower records were around $P_{n+1|n}\approx0.45$. More detailed results for this probability in time series of different interval lengths $N$ can be found in Tab.~\ref{Tab:Stock_Rec_corr}. Apparently, the conditional probabilities $P_{n+1|n}$ for upper and the lower records are much smaller than $0.5$ independent of the interval length $N$.

\begin{table}
\begin{center}
\begin{tabular}{|c||c||c|c|c|c|c|}
  \hline
  \quad\quad & Sy. RW. & N=5000 & N=1000 & N=250 & N=100 & N=25 \\
  \hline\hline
  Up. rec. & 0.5 & 0.440 & 0.444 & 0.448 & 0.443 & 0.451 \\
  Lo. rec. & 0.5 & 0.460 & 0.459 & 0.452 & 0.450 & 0.444 \\
  \hline
\end{tabular}
\end{center}
 \caption{ \label{Tab:Stock_Rec_corr} Correlations $P_{n+1|n}$ between records in successive trading days in daily stock data from the S\&P 500 for different interval length $n=5000,1000,250,100,25$ compared with the result of $P_{n+1|n} = 1/2$ for the symmetric random walk. }
\end{table}

When we compare these findings with numerical simulations of $P_{n+1|n}$ for the AR(1) process, we find a better agreement. When simulating a $n=5000$ step symmetric AR(1) process with $\alpha=0.99$ (as in Fig.~\ref{Fig:ar1_250}), we find a value of $P_{n+1|n} \approx 0.4583...$ . A small bias of the order $c=0.02$, which was found in the data, had only a weak effect, smaller than $0.01$, on this probability. From this point of view, the autoregressive AR(1) process with a manually fitted $\alpha$ models the record statistics of the daily stocks more precisely, but again the actual $\alpha$ obtained from the data is too small to describe the behavior of  $P_{n+1|n}$ accurately. 

Looking at the GARCH(1,1) process, we find no deviations in the inter-record correlations from the GRM. For a GARCH(1,1) with a Students-T return distribution and the maximum-likelihood parameters ($\alpha_0=0.000019,\alpha_1=0.12$ and $\beta_1=0.85$) we find that to a very high accuracy $P_{n+1|n} = 1/2$. Consequently, the deviations in the correlations are presumably generated by the autoregressive behavior of the stock prices and not by the fluctuations of the standard deviation. For the combined autoregressive GARCH(1,1) model we find inter-record correlations with $P_{n+1|n} \approx 0.4748...$, which agree only moderately with the observed correlations. It seems that the combined model, which gives an accurate description of the mean record number $\langle R_n\rangle$ of the stock prices, does not fully capture the nearest-neighbor correlations between individual stock records. At this point, the reason for this discrepancy remains unclear to the author.

\subsubsection{Full distribution of the record number}

\begin{figure*}
\includegraphics[width=0.48\textwidth]{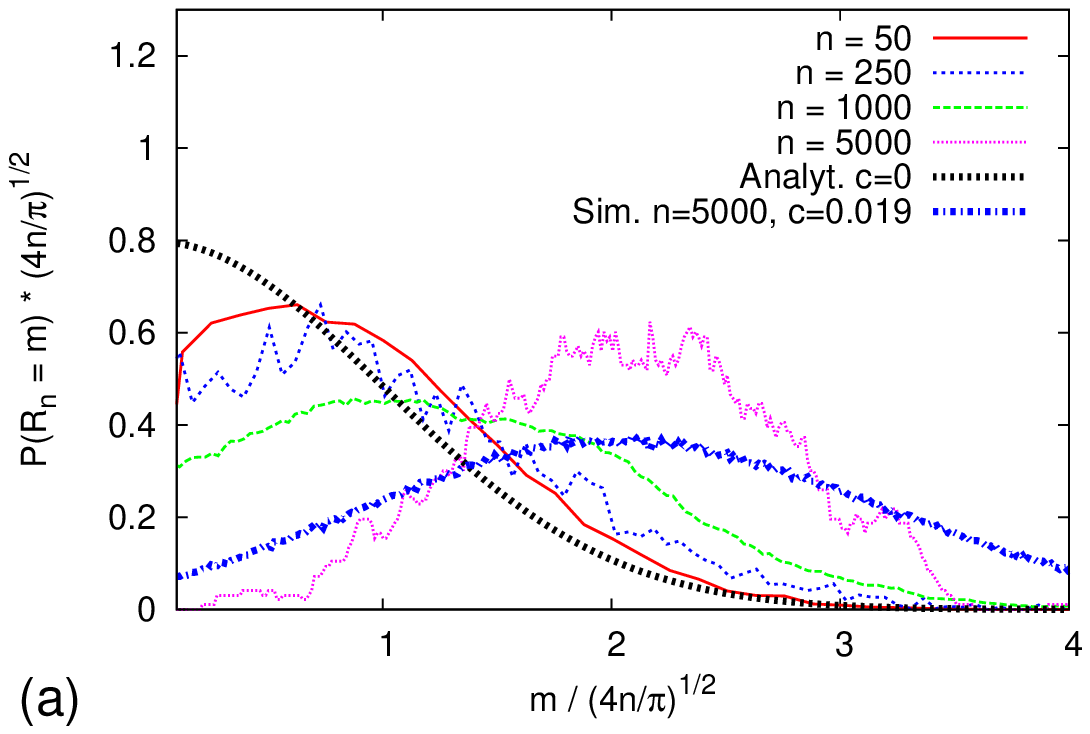}
\includegraphics[width=0.48\textwidth]{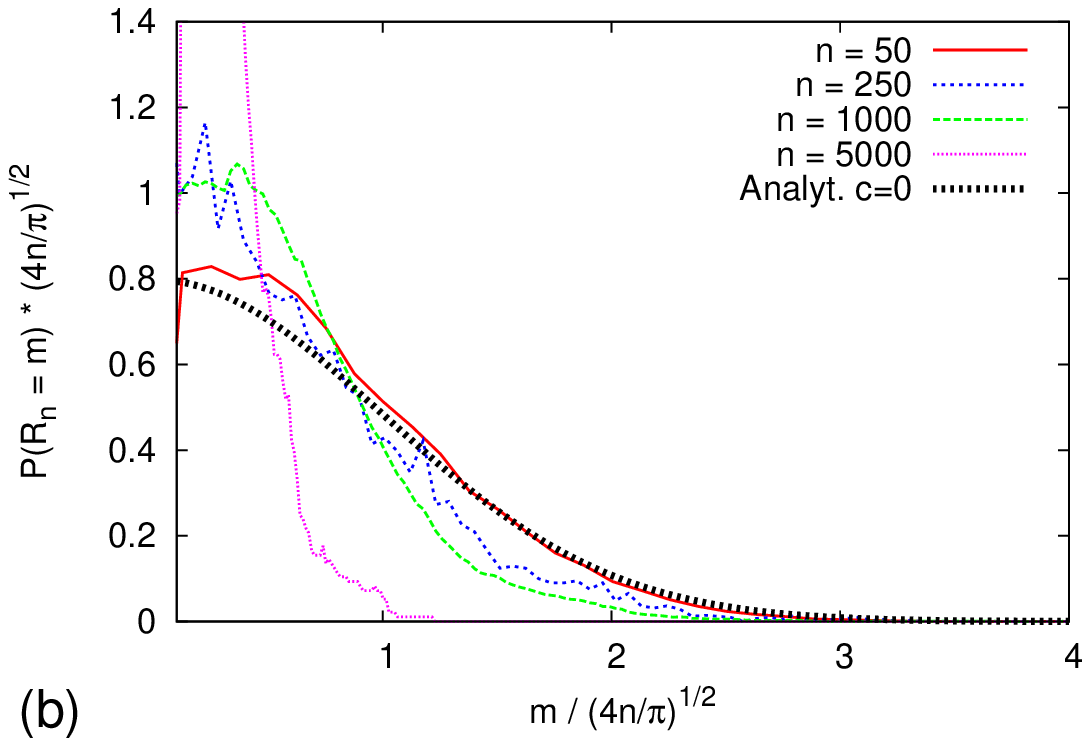}
\caption{\label{Fig:R_n_distr_rescaled} (Color online) (a) The rescaled upper record number of the daily stock prices $S_n$ of the S\&P 500 index. We computed the upper record number at the end of intervals with different interval lengths of $n=50,250,1000,5000$ trading days. For the intervals shorter than $n=5000$ we averaged the distributions over consecutive intervals. All distributions are normalized and rescaled with respect to the mean record number $\langle R_n^{\left(\textrm{GRM}\right)}\rangle = \sqrt{4n/\pi}$ of the symmetric random walk. The distribution of the symmetric random walk as well as a numerical distribution for a biased random walk ($c=0.019$) after $n=5000$ time steps is also plotted for comparison. (b) The same analysis for the lower record number of the daily stock prices. }
\end{figure*}

To conclude our study of the record statistics of the stock prices $S_n$, we also analyzed the full distribution of the record number. Again, we tried to compare the distribution of the record number in the stocks with the predictions from the GRM. As shown in \cite{Majumdar2008}, for an unbiased random walk the record number $R_n$ has a half-Gaussian distribution. For $n\rightarrow\infty$, we have
\begin{eqnarray}
 \textrm{Prob}\left[R_n = m\right] \approx \frac{1}{\sqrt{n\pi}} \textrm{exp}\left(-\frac{m^2}{4n}\right),
\end{eqnarray}
for a positive integer $m\in\left[1,n\right]$. Therefore, the rescaled record number $R_n / \langle R_n\rangle$ is distributed according to a half Gaussian Standard-Normal distribution. Interestingly, the most probable record number in the unbiased case is always $R_n=1$. In \cite{Majumdar2012}, the asymptotic behavior of the distribution $P\left[R_n = m\right]$ was studied also for the biased case. It was shown that for $n\rightarrow\infty$ the record number of a Gaussian random walk (L\'evy index $\mu=2$) approaches a Gaussian distribution. Unfortunately, the results presented in this publication do not hold in the regime of $c\sqrt{n}\ll\sigma$, which is most interesting for our analysis of stock data.

\begin{figure}
\includegraphics[width=0.48\textwidth]{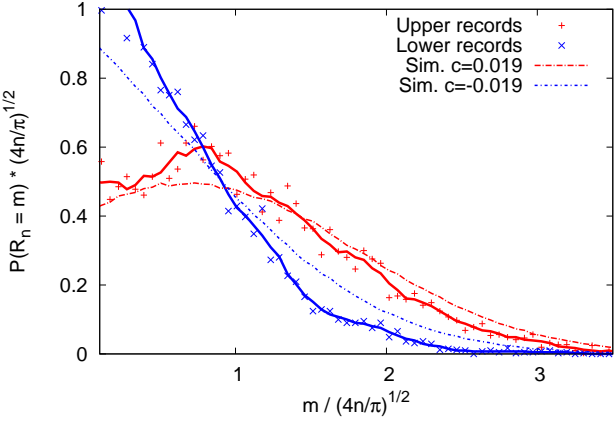}
\caption{\label{Fig:R_n250_distr} (Color online) The rescaled upper and lower record number of the daily stock prices $S_n$ of the S\&P 500 index. Here, we computed the upper and lower record number at the end of 20 intervals with 250 trading days length. Again, these distributions are normalized and rescaled with respect to the mean record number of the symmetric random walk with $\langle R_n^{\left(GRM\right)}\rangle = \sqrt{4n/\pi}$. The blue and red dashed lines are results from numerical simulations of biased random walks with a Gaussian jump distribution ($\sigma=1$) and a drift of $c=0.019$. The drift was obtained from the S\&P 500 data by linear regression in the intervals.}
\end{figure}

To illustrate how the distribution of the record number of the stocks in the S\&P 500 evolves in time, we computed this distribution for different interval length $n$. In Fig.~\ref{Fig:R_n_distr_rescaled}, we show rescaled distributions of the mean record number divided by $\langle R_n^{\left(GRM\right)}\rangle = \sqrt{4n/\pi}$ at the end of intervals with length $n$. As in our previous considerations, if possible, we averaged over successive intervals. 

The left panel in Fig.~\ref{Fig:R_n_distr_rescaled} shows distributions of upper records of $S_n$. Here, for small interval length the distribution of the record number $R_n$ is still very similar to the analytical prediction for the unbiased random walk, but already for $n=50$ the most probable record number is not one anymore and the maximum of the distribution is shifted. For $n=5000$ the distribution resembles a full Gaussian and appears to be symmetric around its mean value.

The right panel in Fig.~\ref{Fig:R_n_distr_rescaled} shows rescaled distributions of lower records. Here, the distribution after $n=50$ trading days is again very similar to the half-Gaussian. With increasing $n$ the distributions get narrower and small record numbers become more and more likely.

In Fig.~\ref{Fig:R_n250_distr}, we compare the rescaled distribution of the upper and lower record number $R_{250}$ after $n=250$ trading days with numerical simulations of biased random walks. We simulated a Gaussian GRM with a jump distribution of $\sigma=1$ and the same linear drift of $c=0.019$ as before. Both the distributions of the upper and the lower mean record number agree relatively well with the model prediction, but again there are some essential deviations between the model and the data. Both in the case of the upper records and in the case of the lower records, the number of series with a small record number is larger than predicted and in both cases the tails of the distributions decay faster. 

\subsection{First-passage times of the stocks}

\begin{figure}
\includegraphics[width=0.48\textwidth]{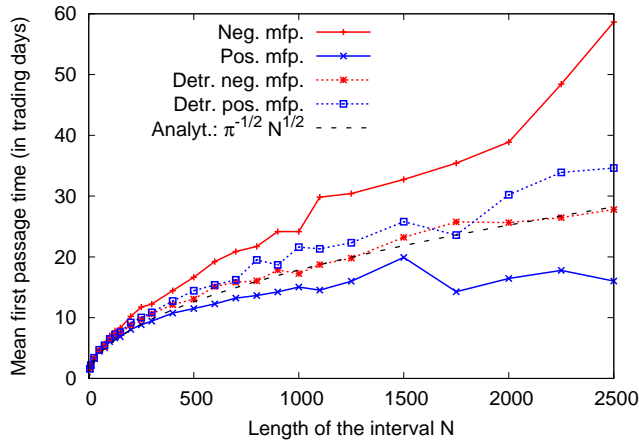}
\caption{\label{Fig:fpt_varN} (Color online) Averaged mean first-passage times (fpt's) for the logarithmic stock prices in the S\&P 500 index. The plot shows both the mean negative (red) and positive (blue) fpt's of the daily stock prices for different interval length $N$. Only the first-passage times smaller than the respective interval length $N$ contributed to the individual data points. We computed the fpt's for the undetrended (bold lines) and the lin-log-detrended (dashed lines) stock prices. The analytical result for the symmetric random walk is given by the black dashed line. }
\end{figure}

To improve our understanding of the record statistics of the S\&P 500 index, we also considered the mean first-passage times (fpt's) $f_{\pm}\left(n\right)$ of the logarithmic stock prices. Since a process with $R_n$ records can be seen as a chain of $R_{n-1}$ first-passage problems (and one survival problem), we assume that we can learn something from considering these quantities. 

Due to the finite length of the time series, it is of course impossible to compute the full mean fpt including first-passage events with arbitrarily large $n>5000$. Because of that, we considered the averaged fpt's on shorter intervals of several different interval lengths $N$. For each entry $S_n$ in a given time series, we computed the time that it took until the logarithmic stock prize $\ln S_n/S_{0}$ of this trading day was first deceeded (negative fpt) or exceeded (positive fpt) by a succeeding logarithmic prize. Stocks that did not cross this initial value within the next $N$ steps were not considered. Then we averaged these fpt's of over all entries and all stocks in an interval, to get the positive and negative mean fpt's for the individual intervals. Eventually, those mean values were averaged over all intervals. We performed this analysis both for the undetrended logarithmic stock prices and the lin-log-detrended stocks, as well as for numerous different choices for the interval length $N$. The results are summarized in Fig.~\ref{Fig:fpt_varN}.

In this figure, we also plotted the analytical result for the symmetric random walk with $f\left(N\right) = \sqrt{N/\pi}$. The results for the undetrended mean fpt's are not surprising. The negative mean fpt is significantly increased in good agreement with the decreased number of lower records in the undetrended data (see also \cite{Wergen2011b}). Of course, if it takes longer until the next negative first-passage event occurs after a lower record, the lower record rate is smaller. Accordingly also the fact that the positive fpt is decreased agrees well with the large number of upper records in the undetrended S\&P 500 data.

If we detrend the data, the mean fpt's of the stocks shift much closer to the analytical result for the symmetric random walk. Nevertheless, our analysis indicates a small asymmetry, which might be related to the slightly decreased number of lower records in the detrended data discussed in the previous section. While the detrended positive fpt's behave exactly as predicted by the random walk model, the negative mean fpt's are increased for most interval lengths $N>500$. 

In future research it might be interesting to have a more detailed look at the fpt's of more sophisticated models like the AR(1) or the GARCH(1,1) with heavy-tailed increments. Preliminary studies indicate that in these models the mean fpt's are increased in consistence with their record statistics.   

\section{Record statistics of multiple stocks}
\label{fin:n_stocks}

\begin{figure*}[t]
\includegraphics[width=0.48\textwidth]{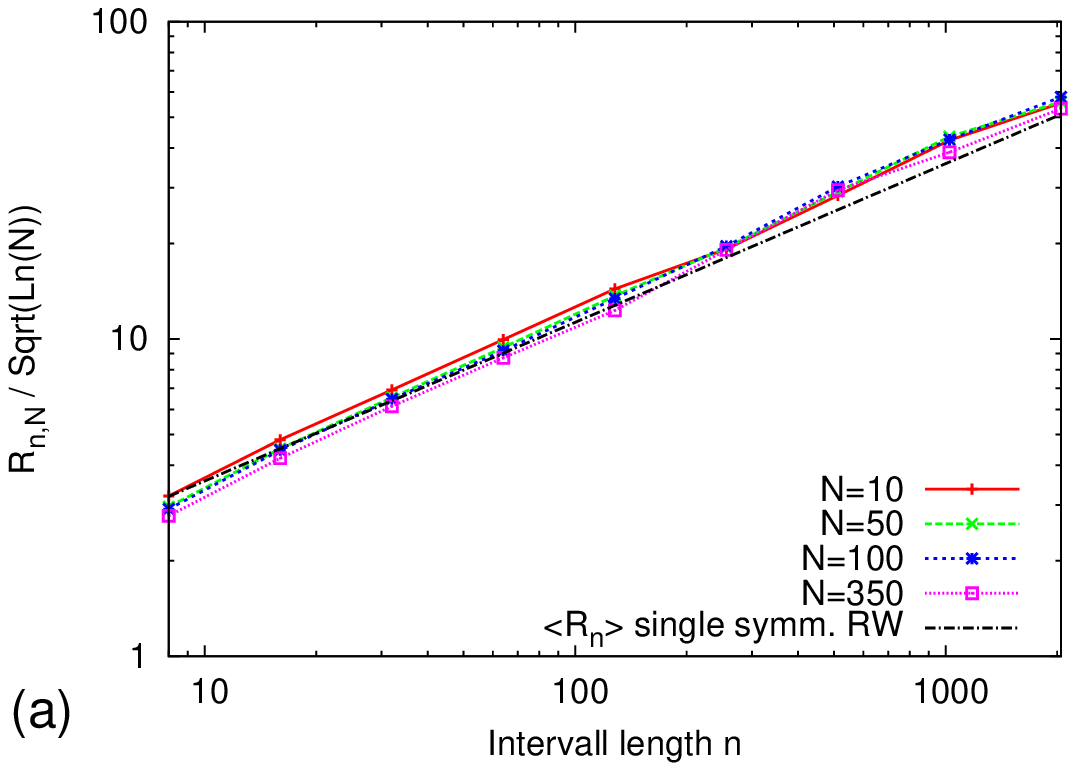}
\includegraphics[width=0.48\textwidth]{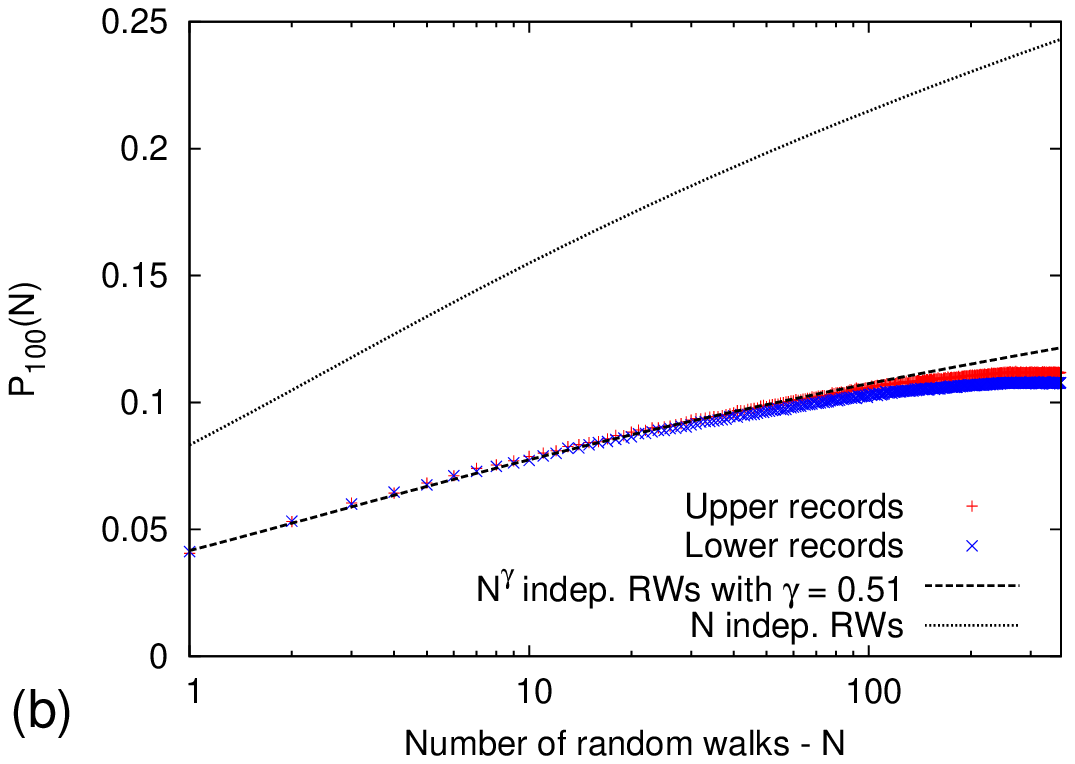}
\caption{\label{Fig:nstocks} (Color online) (a) $\langle R_{n,N}\rangle / \sqrt{\ln N}$ of the maximum of $N$ randomly selected detrended and normalized stocks from the S\&P 500. The $5000$ trading days were subdivided in intervals of $n$ trading days and then linearly detrended in these intervals using the average linear trend of the index. The jump distributions were rescaled and normalized to standard deviation unity. Then, we chose $N$ stocks randomly out of the total number of $366$ stocks and analyzed the evolution of the record number in this set. We computed $\langle R_{n,N}\rangle/ \sqrt{\ln N}$ at the end of the intervals and averaged over $k$ subsequent intervals with $nk\leq5000$. We averaged over $10^4$ randomly selected ensembles for each $N$. (b) The averaged record rate $P_{n,N}$ after $n=100$ trading days. The analysis was performed as before. The dashed line gives our analytical prediction for $N$ Gaussian random walks multiplied with a manually fitted prefactor of $\gamma=0.51$. The bold black line is the analytical prediction for $N$ independent random walks (see \cite{Wergen2012}).}
\end{figure*}

In a recent article of Wergen et al.~\cite{Wergen2012}, the record statistics of ensembles of $N$ independent and symmetric random walks was studied. In that publication, it was shown that the mean record number $\langle R_{n,N}\rangle$ of the maximum of $N\gg1$ independent Brownian random walks with a common jump distribution that has a finite variance ($\mu=2$), is given by
\begin{eqnarray}
 \langle R_{n,N} \rangle \approx 2 \sqrt{n\ln N}.
\end{eqnarray}
The record rate $P_{n,N}$ of the maximum of these $N$ Brownian walkers behaves like
\begin{eqnarray}\label{fin:P_nN}
 P_{n,N} \approx \sqrt{\frac{\ln N}{n}}.
\end{eqnarray}

Furthermore, one can compute the full distribution of the record number $R_{n,N}$ in this case, which approaches a Gumbel form \cite{Majumdar2012,Abramowitz1970}. Wergen et al.~also considered the L\'evy-regime with $\mu<2$, where, surprisingly, $\langle R_{n,N}\rangle$ becomes completely independent of $N$ for $N\gg1$.

In \cite{Wergen2012}, the results for the Brownian case were compared to stock data from the S\&P 500 index. To make this comparison, one can consider randomly chosen subsets of size $N$ of the $366$ stocks and analyze the record statistics of their maximum. To make these $N$ stocks comparable, it is necessary to detrend and rescale the individual time series. In \cite{Wergen2012}, the logarithms $\ln S_n/S_0$ of the stocks were first detrended with respect to the index mean value, before they were normalized in a way so that the standard deviation of the jump distribution was one. This way, one compares $N$ detrended and rescaled stocks with a common variance, similar to $N$ symmetric random walks with a Gaussian Standard Normal jump distribution.

The left panel of Fig.~\ref{Fig:nstocks} shows a plot of the ratio $\langle R_{n,N}\rangle / \sqrt{\ln N}$ against the interval length $n$. The mean record number $\langle R_{n,N}\rangle$ of the maximum of $N$ index-detrended and rescaled stocks at the end of time series with length $n$ was computed for different values of the ensemble size $N$. For each $N$, we averaged over many different randomly selected subsets. For a given interval length $n$, we considered $\langle R_{n,N}\rangle$ on as many consecutive intervals as we could obtain from the total number of $5000$ trading days. 

Apparently, the curves for all considered values of $N$ collapse and we find that $\langle R_{n,N}\rangle / \sqrt{\ln N}$ is independent of $N$. This confirms that the record statistics of the maximum of $N$ index-detrended and normalized stocks from the S\&P 500 has the same dependence on $N$ as the maximum of an ensemble of $N$ independent random walks. In other words, we find that, in both cases
\begin{eqnarray}
 \langle R_{n,N}\rangle \propto \sqrt{n\ln N}.
\end{eqnarray}

In the right panel of Fig.~\ref{Fig:nstocks}, we present an analysis of the record rate $P_{n,N}$ of $N$ randomly selected, index-detrended and normalized stocks compared with the analytical prediction from \cite{Wergen2012} (Eq.~\ref{fin:P_nN}). The record rate was evaluated at the end of intervals with a fixed length of $n=100$. This figure is consistent with the findings in \cite{Wergen2012}. Both the record rates of the stocks and of the $N$ independent random walkers are proportional to $\sqrt{\ln N}$ and we have
\begin{eqnarray}
 P_{n,N} \propto \sqrt{\frac{\ln N}{n}},
\end{eqnarray}
but the prefactors are different. In this case, for $n=100$, the record rate of the stocks is only $0.51$ times the record rate of the random walkers. 

In \cite{Wergen2012}, a similar prefactor was found considering the mean record number on intervals of $n=250$ trading days length. A possible way to explain this prefactor is the following: As it is well known, while the theory in \cite{Wergen2012} was developed for $N$ entirely independent random walkers, the stocks in the S\&P 500 are strongly correlated. Therefore, the fact that the record number of the maximum of the stock prices has a smaller prefactor, can be explain by assuming that only a smaller number $\tilde{N}<N$ of effectively independent stocks contributes to the record statistics.

The right panel of Fig.~\ref{Fig:nstocks} shows that the record rate of $N$ randomly selected stocks is the same as the record rate of $N^{\gamma}$ independent random walkers (with $\gamma\approx0.51$). Therefore, we can conjecture that only $N^{\gamma}$ stocks are independent in the context of record statistics. This hypothesis is supported by the fact that the record rate of the stocks saturates at a constant value for $N>100$. Apparently, here, the number of effectively independent stocks saturates at an upper limit of $366^{\gamma}$. 

\section{Weekly distributions of records}
\label{fin:weekly_monthly}

Since not all of the findings for the record statistics of the stock data are in perfect agreement with the predictions of our simple stochastic models, we further investigated the specific patterns of record occurrence in the stock data. One thing that we were interested in is the weekly distribution of record-breaking events. The simple question was: Does the number of upper and lower records in the stocks depend on the weekday?

\begin{figure}[t]
\includegraphics[width=0.48\textwidth]{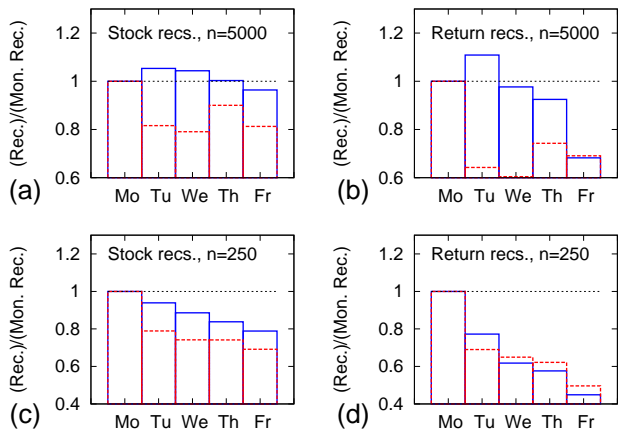}
\caption{\label{Fig:weekday_records} (Color online) Weekly distributions of record-breaking stock prizes and daily returns. We counted the number of records on a given weekday and divided by the number of records on Mondays to make the distributions comparable. (a) Records of the daily stock prizes $\ln S_n/S_0$ in the time series of $n=5000$ trading days. (b) Records of the daily returns $\ln S_n/S_{n-1}$ for the time series of $n=5000$ trading days. (c) Records of the daily stock prizes $\ln S_n/S_0$ in $20$ intervals of each $n=250$ trading days. (d) Records of the daily returns $\ln S_n/S_{n-1}$ in $20$ intervals of each $n=250$ trading days. In all considered cases the highest number of records occurs on Mondays and Tuesdays, then the record rate decreases in the course of the weak. This effect is far more pronounced for the shorter time-interval.}
\end{figure}

The weekly distributions of records in stock prices and stock returns for two different interval length are presented in Fig.~\ref{Fig:weekday_records}. In the upper two figures (a,b) this distribution was computed for records in the entire time series of $5000$ trading days. The upper left panel (a) shows the distribution of the records of the stock prices and the upper right the one (b) of the return records. The number of records that was recorded on a given weekday was divided by the number of records that was set on Mondays to make the plots more comparable. In both cases, records are more likely on Mondays than on Fridays and, with few exceptions, the record rate decreases over the week. 

The picture gets clearer if we consider the weekly distribution of record stock prizes and record returns set on intervals with a shorter length of only 250 trading days. This analysis can be found in the two lower panels of Fig.~\ref{Fig:weekday_records} (c,d). Here, the occurrence of records decreases monotonically over the week. For the return records this effect is a lot stronger and the record rate on Mondays is roughly twice as large as on Fridays.

These significant effects on shorter time-scales show that the simple models introduced above are only capable to predict certain properties of the record statistics of the stock prices. Up to a certain degree, they can model the averaged record rate and the mean record number of the stocks on time-scales much longer than one week. More complicated factors, such as the weekly fluctuations of market activity, are not reproduced. 

\section{Summary and conclusions}
\label{fin:summary}

After briefly recapitulating several important results for the record statistics of time series of independent and identically distributed (i.i.d) random variables (RV's) and symmetric discrete-time random walks in section \ref{fin:iid_rvs}, we discussed the occurrence of record-breaking events in several stochastic processes popular in financial modeling. In section \ref{fin:random_walks} we described various asymptotically exact results for the record statistics of discrete-time random walks with a bias. Furthermore we presented some findings for the record rate and the mean record number of a biased random walk in the regime of small drift $c$ and relatively small series length $n$ (with $c\sqrt{n}\ll\sigma$). Here, the effect of the bias on the record rate is particularly simple and linear in $c$.

Following this, in section \ref{fin:autoregressive}, we introduced two important and more complicated stochastic processes. We performed a detailed numerical analysis of the record rate of the autoregressive AR(1) process. For $\alpha<1$ and $n\rightarrow\infty$ the record rate of this process converges to the i.i.d. result with $p_n=1/n$. For small values of $n$ and $\alpha$ close to one, the record statistics of the AR(1) process is similar to the one of the symmetric random walk. Our numerical simulations indicated that for very small $1-\alpha\ll1$, the record rate $P_n^{\left(\alpha\right)}$ decays exponentially in $1-\alpha$.

In addition to the AR(1) process, we considered the GARCH(1,1) model with a second stochastic process describing a time-dependent standard deviation of the increments. Interestingly, it seems that the occurrence of records in symmetric and stationary Gaussian GARCH(1,1) models is entirely identical to the occurrence of records in the symmetric random walk, independent from the choice of the parameters of the model. In contrast to the symmetric random walk however, the record statistics of the GARCH(1,1) process depends crucially on the shape and the tail-behavior of the underlying jump distribution. With an application on stock data in mind, we introduced the symmetric Students-T distribution that decays like $1/|\xi|^{\nu}$ for $|\xi|\rightarrow\infty$. We demonstrated numerically that the record rate and the mean record number of the GARCH(1,1) decreases with decreasing $\nu$. For very heavy-tailed distributions the occurrence of records is significantly reduced in comparison to the symmetric random walk.

In sections \ref{fin:single_stocks}, \ref{fin:n_stocks} and \ref{fin:weekly_monthly} we presented a detailed analysis of the statistics of record-breaking events in time series of daily stock prizes. Both the record process of the daily returns $\delta_n$ and the record process of the daily stock prizes $S_n$ behave similar to what is predicted for the Geometric Random Walk Model (GRM). In both cases, the agreement gets better if the length of the considered series length decreases. Indeed, for time series length of $n=50$ trading days or smaller, the GRM describes the averaged record statistics of the stocks accurately.

We found that the record rate of the daily returns over the full period of 5000 trading days fluctuates strongly and is dominated by a few periods of very high market activity (or crisis). Additionally, the records of the returns are strongly correlated. Upper return records are often followed by lower return records and vice versa. 

For time series with an intermediate length of $n=250$ trading days, the record statistics of the daily returns is already more similar to the i.i.d.~behavior, but we find that the rate of lower records is slightly increased by a mechanism that we can not explain. For $n=250$ there are more lower than upper return records. For shorter time series of $n=100$ trading days this effect vanishes and the record rate of the returns becomes very similar to the i.i.d.~record rate of $p_n^{\left(\textrm{iid}\right)}=1/n$. 

Our analysis of the stock prizes showed that the GRM slightly overestimates the mean record numbers $\langle R_n\rangle$ on intervals of intermediate length with $100<n<1000$. We demonstrated that an autoregressive AR(1) process with a manually fitted parameter $\alpha=0.99$ models the behavior of the record numbers more accurately. However, the parameter of $\alpha=0.99$ is not in agreement with the value of $\alpha=0.9993$ we obtained from the stock data using maximum-likelihood estimation. For this observational value of $\alpha$, the effect is too small to explain the asymmetry in the mean record number. 

Following this we compared the stock data to the more sophisticated GARCH(1,1) process. Such a model with heavy-tailed, Students-T distributed increments can also explain a decreased mean record number $\langle R_n\rangle$ and the observed asymmetry. Yet again, the effect is too weak to explain the behavior of the actual stock data. In this context we also compared the stock data to a combined autoregressive GARCH(1,1) model with Students-T returns, which can effectively model the behavior of the mean record number of the stock data to a good accuracy without any manually fitted parameters. Therefore, we assume that the interesting asymmetry in the mean record number is generated by a combination of both the autoregressive and the GARCH-like behavior of the stocks.

The AR(1) process is also better in describing the inter-record correlations of the stock prices, which are significantly weaker than in the random walk case. We considered the full distribution of the mean record number of the stock prices, which are, as expected, slightly narrower than the corresponding distributions from the GRM. These findings support the hypothesis that the stock prizes are more accurately described by an autoregressive process, which is basically like a random walker in a quadratic potential that draws the walker back to some mean value. It would be nice to compute the record statistics of such an AR(1) process analytically. Such a result might eventually lead to a better understanding of record-breaking stock prices.

As in \cite{Wergen2012a}, we discussed the record statistics of the maximum of multiple stocks that were detrended and normalized. Our analysis in section \ref{fin:n_stocks} showed that the mean record number $\langle R_{n,N}\rangle$ of the maximum of $N$ of these stocks has the same dependence on $N$ as the maximum of $N$ independent Gaussian random walks. Nevertheless, the mean record number of the stocks had a different prefactor that can be explained by introducing an effective number $N^{\gamma}$ of independent stocks. We assume that only these $N^{\gamma}$ stocks are effectively uncorrelated. It would be interesting to see if this effective number of stocks and the coefficient $\gamma$ can also be measured by other methods unrelated to record statistics. By now, we are not aware of similar results for stock data, even though such a measure of correlation between the stocks might be useful in the estimation of possible risks for instance for stock portfolios and index funds.

In the penultimate section \ref{fin:weekly_monthly}, we also studied the weekly distribution of the occurrence of record events both in stock prizes and in daily returns. We showed that record events are most likely to occur in the beginning of the week. This is not surprising, since it is known that the market activity varies over the week and usually decreases when the weekend approaches \cite{Chordia2002,Dacorogna1993}. 

In summary, we demonstrated, where the simple Geometric Random Walk Model is useful in describing the record statistics of stock prizes from the S\&P 500 stock market and we illustrated where it fails. With the slightly more complicated AR(1) and the GARCH(1,1) process we already gave ideas how to improve the modeling. With a combined autoregressive GARCH(1,1) model with a heavy-tailed return distribution we were able to reproduce the behavior of the mean record number of the stock data to a good accuracy. However, a model that describes the strong fluctuations of the record rate discussed in this article and effects like the distinct weekday-dependence of the record occurrence is still missing. Such a model would be an interesting goal for future research. Similarly the nontrivial dependence of the ensemble mean record number $\langle R_{n,N}\rangle$ of the stocks on $N$ certainly deserves a better explanation.

\smallskip
\textbf{Acknowledgements} --- The author is grateful to Joachim Krug for many interesting and insightful discussions on the subject of this article. The analysis of the weekly distribution of the records was motivated by Sid Redner. Furthermore, the author would like to thank Ivan Szendro, Jasper Franke, Satya Majumdar, Gr\'egory Schehr, Bernd Rosenow, Rudi Sch\"afer and Thomas Guhr for their input and their suggestions. 

\bibliographystyle{apsrev4-1}
\bibliography{lit_finance_pre}{}

\end{document}